\documentclass[twocolumn]{aastex631}

\usepackage{CJKutf8}

\usepackage{float}
\usepackage{multirow}
\usepackage{subfigure}
\usepackage{booktabs}
\usepackage{amsmath}
\usepackage{graphicx}
\usepackage[section]{placeins}

\begin{document}

\title{Neutrino constraints and detection prospects from gamma-ray bursts with different jet compositions}

\correspondingauthor{Hou-Jun L\"{u} }
\email{lhj@gxu.edu.cn}

\author[0009-0007-5062-3221]{Yang-Dong-Jun Ou}
\affiliation{Guangxi Key Laboratory for Relativistic Astrophysics, School of Physical Science and Technology, Guangxi University, Nanning 530004, People's Republic of China}

\author[0000-0001-6396-9386]{Hou-Jun L\"{u} }
\affiliation{Guangxi Key Laboratory for Relativistic Astrophysics, School of Physical Science and Technology, Guangxi University, Nanning 530004, People's Republic of China}

\author{Xue-Zhao Chang }
\affiliation{Guangxi Key Laboratory for Relativistic Astrophysics, School of Physical Science and Technology, Guangxi University, Nanning 530004, People's Republic of China}

\author{Xiao-Xuan Liu }
\affiliation{Guangxi Key Laboratory for Relativistic Astrophysics, School of Physical Science and Technology, Guangxi University, Nanning 530004, People's Republic of China}

\author[0000-0002-7044-733X]{En-Wei Liang }
\affiliation{Guangxi Key Laboratory for Relativistic Astrophysics, School of Physical Science and Technology, Guangxi University, Nanning 530004, People's Republic of China}

\begin{abstract}
The prompt emission mechanism of gamma-ray bursts (GRBs) is a long-standing open question, and GRBs have been considered as potential sources of high-energy neutrinos. Despite many years of search for the neutrino events associated with GRBs from IceCube, there were no results. However, the absence of search results for neutrinos provides a unique opportunity to constrain the parameter space of GRB jet models. In this paper, we chose four peculiar GRBs with two different types of jet composition to investigate neutrino emission. It is found that only GRB 211211A could be well constrained within the dissipative photosphere model. By adopting the specific parameters of the photosphere, one can obtain \(\varepsilon _{p } \text{/} \varepsilon _{e }<8\) for \(f_{p}>0.2\) from GRB 211211A. For the Internal-collision-induced Magnetic Reconnection and Turbulence (ICMART) model, we can effectively constrain neither GRB 230307A nor GRB 080916C. Moreover, we also investigate the detection prospects of high-energy neutrinos from GRBs and find that it is difficult to detect at least one high-energy neutrino associated with GRBs from the ICMART model even during the IceCube-Gen2 operation. For the GRB 211211A-like events, it is possible to detect at least one neutrino coincident with the gravitational wave during the IceCube-Gen2 operation, if such an event originated from mergers of compact stars within the photosphere dissipation.

\end{abstract}


\begin{CJK*}{UTF8}{gbsn}
\section{Introduction}\label{sec:1}
\end{CJK*}

Gamma-ray bursts (GRBs), the most energetic phenomena at cosmological distance, have been proposed to be the candidates for ultrahigh-energy cosmic-ray (UHECR) accelerators \citep[e.g.,][]{1995ApJ...449L..37M,1995ApJ...453..883V,1995PhRvL..75..386W,2008PhRvD..78b3005M,2015APh....62...66B} and the potential emitters for extragalactic neutrinos \citep[e.g.,][]{1997PhRvL..78.2292W,2004APh....20..429G,2006PhRvD..73f3002M,2006PhRvL..97e1101M,2006ApJ...651L...5M,2012PhRvL.108w1101H,2012JCAP...11..058G,2013PhRvL.110l1101Z,2016PhRvD..93l3004L,2017ApJ...848L...4K,2021JCAP...05..034P,2023MNRAS.520.2806C,2023ApJ...950..190M,2024ApJ...974..185M}. MeV neutrinos can be produced through several mechanisms in the central engine and the outflow. For example, the neutrino-dominated accretion flow \citep{2016PhRvD..93l3004L,2022ApJ...925...43Q}, or $r$-process in mergers of binary neutron stars \citep{2023MNRAS.520.2806C}. However, such MeV neutrinos are very difficult to detect except for the location of GRBs close enough within a few hundred kiloparsecs \citep{2016PhRvD..93l3004L,2022ApJ...925...43Q,2023MNRAS.520.2806C}. High-energy neutrinos with an energy range from TeV to EeV, on the other hand, can be produced in both the internal dissipation of prompt emission and the external shock of afterglow (\citet{2022arXiv220206480K} for a review). The current neutrino detectors, such as IceCube \citep{2017JInst..12P3012A} and ANTARES \citep{2011NIMPA.656...11A}, are used to detect high-energy neutrinos. Moreover, several near-future neutrino detectors, like Baikal-GVD \citep{2011NIMPA.639...30A}, IceCube-Gen2 \citep{2021JPhG...48f0501A}, KM3NeT \citep{2016JPhG...43h4001A}, P-One \citep{2020NatAs...4..913A} and TRIDENT \citep{2023NatAs...7.1497Y}, are also proposed or under construction, and detect the high-energy neutrino in the future.

Since the discovery of GRBs about half a century ago, the mechanism of GRB prompt emission remains under debate \citep{2018pgrb.book.....Z}. In general, two scenarios of jet composition, a matter-dominated fireball and a Poynting-flux-dominated outflow, have been proposed to interpret the observations of GRBs \citep{2018pgrb.book.....Z}. For the matter-dominated fireball, the emission site would be the photosphere radius \(R_{\text{ph}}\sim 10^{10}–10^{12} \text{cm}\) \citep{2005ApJ...628..847R}, or the internal shock radius \(R_{\text{IS}}\sim 10^{13}–10^{14} \text{cm}\) \citep{1994ApJ...430L..93R}. For the Poynting-flux-dominated outflow, one representative magnetic dissipation model is the Internal-collision-induced Magnetic Reconnection and Turbulence (ICMART) model \citep{2011ApJ...726...90Z}. The ICMART model invokes the collision of Poynting-flux-dominated shells at large radii and then results in magnetic dissipation through magnetic reconnection. Sometimes, the radiations from the photosphere and internal shock would be strongly suppressed by a larger magnetization parameter which is defined as the flux ratio between Poynting-flux and matter outflows. Moreover, the existence of mini-jets \citep{2014ApJ...782...92Z} and large prompt emission radius \(R_{\text{ICMART}}\sim 10^{15}–10^{16} \text{cm}\) are predicted by the ICMART model. On the other hand, the predicted detection of neutrino flux from different GRB jet models is quite different \citep{2013PhRvL.110l1101Z}.

Over 15 yr of operation, the IceCube Collaboration has performed dedicated analyses to search for high-energy neutrino events associated with GRBs, but nondetection of any neutrino events has been associated with GRBs \citep{2022ApJ...939..116A}. However, the nondetection of any neutrinos can not only place a stringent upper limit for the time-integrated muon-neutrino flux, but also provides significant constraints on the parameters space of GRB jet models, such as emission radii and bulk motion Lorentz factor \citep[e.g.,][]{2012ApJ...752...29H,2012Natur.484..351I,2012PhRvD..85b7301L,2013ApJ...772L...4G,2015ApJ...805L...5A,2016ApJ...824..115A,2017ApJ...843..112A,2022ApJ...941L..10M,2023ApJ...944..115A}.

From the observational point of view, several lines of observational evidence suggest that a purely quasi-thermal component in GRB 090902B and a purely nonthermal emission from the magnetic dissipation in GRB 080916C were observed by Fermi Gamma-ray Burst Monitor 
\citep[GBM;][]{2009Sci...323.1688A,2009ApJ...700L..65Z,2010ApJ...709L.172R,2011ApJ...726...90Z}. More recently, several peculiar GRBs, such as the brightest-of-all-time (BOAT) GRB 221009A \citep{2023ApJ...946L..24W,2023ApJ...946L..31B}, the short GRB with extended emission GRB 211227A \citep{2022ApJ...931L..23L}, and a short GRB with extended emission associated with kilonova, GRB 211211A \citep{2022Natur.612..223R,2022Natur.612..228T,2022Natur.612..232Y,2023ApJ...943..146C,2023NatAs...7...67G} and GRB 230307A \citep{2023arXiv230705689S,2023arXiv231007205Y,2024ApJ...962L..27D,2024MNRAS.529L..67D}, have attracted great attention. Moreover, the jet composition of GRBs 211211A and 230307A which are possible originated from merger of compact stars are quite different \citep{2023ApJ...943..146C,2023arXiv231007205Y,2024ApJ...962L..27D,2024MNRAS.529L..67D}. The discovery of the mini-jet of GRB 230307A is consistent with the prediction of the ICMART model \citep{2023arXiv231007205Y}, suggesting that the jet composition of GRB 230307A is magnetically dominated, with a prompt emission radius of $\sim 10^{16}~\rm cm$ \citep{2024ApJ...962L..27D,2024MNRAS.529L..67D}. On the contrary, the presence of thermal emission of GRB 211211A \citep{2023ApJ...943..146C} suggests that the magnetization parameter at the photosphere is not sufficiently large to suppress photosphere emission. The radiation of prompt emission may be from the photosphere radius or the internal shock radius.

On the other hand, there is a nondetection of high-energy neutrinos associated with GRB 221009A \citep{2023ApJ...946L..26A} and GRB 230307A \citep{2023GCN.33430....1I} from observations. Adopting this nondetection, model predictions of high-energy neutrinos can be used to constrain the jet composition and physical parameters of GRB 221009A \citep{2022ApJ...941L..10M,2023ApJ...944..115A} and GRB 230307A \citep{2023ApJ...958..113A}. In this paper, we present the general formula for the calculation of the high-energy neutrino spectrum of GRBs in Section \ref{sec:2}; In Section \ref{sec:3}, we show the constraints on the physical parameters from four GRBs (e.g., 080916C, 090902B, 211211A, and 230307A) that have different jet compositions. The neutrino-detected prospects from the above four GRBs with different jet composition are shown in Section \ref{sec:4}; Finally, the conclusion and discussion are presented in Section \ref{sec:5}. Throughout this paper, we adopt the convention \(Q_{x} = Q/10^{x}\) in CGS units, and use the symbol \(Q^{'}\) to denote quantities in the comoving frame. The cosmological parameters with Hubble constant \(H_{0}=70 
\text{ km s}^{-1}\text{ Mpc}^{-1}\),  \(\Omega_{m}=0.3\), and \(\Omega_{\Lambda}=0.7\), are adopted to do the calculations.

\section{High-energy Neutrino Emission from GRBs}
\label{sec:2}

The primary channel to produce the high-energy neutrinos from GRBs is the hadronic process \citep{2018pgrb.book.....Z}. Both electrons and protons are accelerated by the internal dissipated energy of the GRB jet. The accelerated electrons via the radiation mechanisms (e.g., synchrotron radiation,  synchrotron self-Compton, and Compton upscattering of a thermal seed photon source) can lead to producing gamma-ray emissions which are observed as the prompt emission of GRBs. The accelerated protons, on the other hand, can interact with the gamma-ray photons, and lead to the production of high-energy neutrinos through the photomeson (\(p\gamma \)) interaction \citep{2013PhRvL.110l1101Z}. Moreover, previous studies have shown that sub-TeV neutrinos can be produced through proton-neutron (pn) decoupling and collisions \citep[e.g.,][]{2000PhRvL..85.1362B,2013PhRvL.111m1102M,2022ApJ...941L..10M,2023ApJ...958..113A,2024ApJ...964..126A}. However, due to the strong atmospheric background in this energy range, the effective area of IceCube in the sub-TeV range is much smaller than that of in the TeV-EeV range \citep[see Figure 8 in][]{2023ApJ...953..160A}. Therefore, we only focus on considering the neutrinos that are from the \(p\gamma \) process and do not consider neutrinos that are from the \(pn\) process.  

Following the method in \cite{2013PhRvL.110l1101Z} and \cite{2023ApJ...944..115A}, the target photon source for photomeson interaction is from the prompt emission of GRB itself. The distribution of the target photons can be described by a Band function \citep{1993ApJ...413..281B}:
 \begin{equation}
     \begin{aligned}
         n_{\gamma } \left ( E_{\gamma} \right ) &= \frac{dN_{\gamma } \left ( E_{\gamma} \right )}{dE_{\gamma}}\\&=n _{\gamma,b }
         \begin{cases}
    \varepsilon ^{\alpha } _{\gamma,b }E_{\gamma}^{-\alpha }, & E_{\gamma}< \varepsilon _{\gamma,b }\\
    \varepsilon ^{\beta } _{\gamma,b }E_{\gamma}^{-\beta },   & E_{\gamma}\ge  \varepsilon _{\gamma,b },
    \end{cases}
     \end{aligned}
 \end{equation}
where \(\varepsilon _{\gamma,b } = \frac{E_{\text{peak}}}{2-\alpha } \) is the break energy in the band spectrum, \(n _{\gamma,b }\) is the specific photon number at the break energy \(\varepsilon _{\gamma,b }\). Additionally, \(\alpha\) and \(\beta\) are the low-energy and high-energy photon spectral indices, respectively. 

The neutrino spectrum, on the other hand, can be described as a double break power law \citep{1997PhRvL..78.2292W,2010ApJ...710..346A}:
 \begin{equation}
     \begin{aligned}
        n_{\nu } \left ( E_{\nu } \right ) &= \frac{dN_{\nu } \left ( E_{\nu } \right )}{dE_{\nu }}  \\&=n _{\nu,1 }
        \begin{cases}
    \varepsilon ^{\alpha_{\nu} } _{\nu,1 }E_{\nu }^{-\alpha_{\nu} }, & E_{\nu }< \varepsilon _{\nu,1 },\\
    \varepsilon ^{\beta_{\nu} } _{\nu,1 }E_{\nu }^{-\beta_{\nu} },   & \varepsilon _{\nu,1 }\le E_{\nu }< \varepsilon _{\nu,2 }, \\
    \varepsilon ^{\beta_{\nu} } _{\nu,1 }\varepsilon ^{\gamma_{\nu}-\beta_{\nu} } _{\nu,2 }E_{\nu }^{-\gamma_{\nu} },   &  E_{\nu }\ge \varepsilon _{\nu,2 },
    \end{cases}
     \end{aligned}
 \end{equation}
where \(n _{\nu,1 }\) is the specific neutrino number at the low-energy break \(\varepsilon _{\nu,1 }\). The neutrino spectral indices (e.g., $\alpha_{\nu}$, $\beta_{\nu}$, and $\gamma_{\nu}$) are determined by the photon spectral indices (e.g., $\alpha$ and $\beta$) and the proton spectral index $p$ for a single power-law distribution \(\frac{dN_{p}}{dE_{p}}\propto E_{p}^{-p}\). Based on first-order Fermi acceleration for strong nonrelativistic shocks, the high-energy tail of accelerated particles should have a power-law distribution, and we adopt \(p=2\) throughout this paper \citep[see details of Chapter 4.4 in the book of][]{2018pgrb.book.....Z}. The neutrino spectral indices can be expressed as  
\begin{equation}
     \begin{aligned}
     \alpha_{\nu} = p+1-\beta,     \beta_{\nu} = p+1-\alpha,   \gamma_{\nu} = \beta_{\nu}+2.
     \end{aligned}
\end{equation}
Here, $\alpha_{\nu}$ and $\beta_{\nu}$ are derived by assuming that the neutrino flux is proportional to the proton flux (with index $p$) and the optical depth of \(p\gamma \) process \(\tau _{p\gamma }\propto n_{\gamma}(E_{\gamma})\) (with index $1-\beta$ for below \(\varepsilon _{\nu,1 }\) and $1-\alpha$ for above \(\varepsilon _{\nu,1 }\)). $\gamma_{\nu}$ is derived by solving the continuity equation for charged particles with synchrotron cooling \citep[e.g. the Chapter 5.1.5 in][]{2018pgrb.book.....Z}, and found that the probability of charged pions without cooling obeys a power-law distribution with index 2.

The low-energy break in the neutrino spectrum \(\varepsilon _{\nu,1 }\) is caused by the break in the photon spectrum \(\varepsilon _{\gamma,b }\). We assume that the energy of protons is responsible for $N$ rounds of pion production, the pion production efficiency \(f_{\pi}\) (the fraction of proton energy going into pion production) can be expressed as $f_{\pi}=1-\left ( 1-\left \langle f_{p\to \gamma}  \right \rangle  \right ) ^{ N}$, where \(\left \langle f_{p\to \gamma}\right \rangle = 0.2\) is the fraction of energy transferred from protons to pions. $N$ relates to the \(p\gamma \) optical depth $\tau_{p\gamma}$, and $N \sim \tau_{p\gamma}$ if $\tau_{p\gamma} < 3$. For a greater $\tau_{p\gamma}$, $N\sim \tau_{p\gamma}^2$ is expected \citep[see the footnote 5 in][]{2023ApJ...944..115A}. For $f_{\pi} = 1-\left ( 1-\left \langle f_{p\to \gamma}  \right \rangle  \right ) ^{N} \approx 1 $, it requires $N=25$ with a larger value of $\tau_{p\gamma}$, namely, $N\sim \tau^{2}_{p\gamma}$, and $\tau_{p\gamma}=5$. So , it requires that \(\tau _{p\gamma }\) is less than 5 if $f_{\pi}$ is smaller than one. If \(\tau _{p\gamma } ^{p}\) (the \(p\gamma \) optical depth for peak energy photons interacting with protons) is less than 5, the low-energy break can be expressed as
    \begin{equation}\label{4}
     \begin{aligned}
     \varepsilon _{\nu,1 } &= \varepsilon _{\nu,1 }^{0} \\&= 7.2\times 10^{5}\text{GeV}\left ( \frac{\varepsilon _{\gamma,b}}{\text{MeV}} \right )  ^{-1}
\left ( \frac{\Gamma }{300} \right )^{2} \left ( 1+z \right ) ^{-2},
     \end{aligned}
 \end{equation}
where \(\Gamma\) and $z$ are bulk motion Lorentz factor and redshift, respectively.\footnote{The derivations of the break energy of the neutrino spectrum can be found in the Appendix \ref{App_a}.} If \(\tau _{p\gamma }>5\), one has the pion production efficiency \(f_{\pi} = 1\), and the neutrino flux no longer significantly increases with \(\tau _{p\gamma }\). Therefore, when \(\tau _{p\gamma } ^{p}>5\), the low-energy break should be modified by invoking the \(\tau _{p\gamma } ^{p}\), which can be expressed as
  \begin{equation}
     \begin{aligned}
    \tau _{p\gamma } ^{p} =8.9L_{\rm GRB,52} \left ( \frac{\Gamma }{300} \right )^{-2}R_{13}^{-1}  \left ( \frac{\varepsilon _{\gamma,b}}{\text{MeV}} \right ),
     \end{aligned}
 \end{equation}
where \(L_{\rm GRB}\) is the isotropic luminosity of GRB prompt emission, and \(R\) is the dissipation radius. If the neutrino energy \(E_{\nu }\) is below \(\varepsilon _{\nu,1 }^{0}\), the optical depth of \(p\gamma \) at \(E_{\nu }\) can be calculated as
 \begin{equation}
     \begin{aligned}
   \tau _{p\gamma }\left (E_{\nu} \right ) = \tau _{p\gamma } ^{p}\left ( \frac{E _{\nu}}{\varepsilon _{\nu,1 }}  \right )^{\beta -1}.
     \end{aligned}
 \end{equation}
If \(\tau _{p\gamma }=5\), the typical neutrino energy is given by \(E_{\nu}=\varepsilon _{\nu,1 }^{0}\left ( \frac{\tau _{p\gamma } ^{p}}{5}  \right ) ^{\frac{1}{1-\beta}}\). The neutrino energy should be replaced by \(\varepsilon _{\nu,1 }\) as the low-energy break of the neutrino spectrum when \(\tau _{p\gamma } ^{p}>5\). If this is the case, the low-energy break \(\varepsilon _{\nu,1 }\) can be expressed as
   \begin{equation}\label{7}
     \begin{aligned}
   \varepsilon _{\nu,1 } =  \varepsilon _{\nu,1 }^{0}\left ( \frac{\tau _{p\gamma } ^{p}}{5}  \right ) ^{\frac{1}{1-\beta}}.
     \end{aligned}
 \end{equation}
Together with Eqs.(\ref{4}) and (\ref{7}), the low-energy break of the neutrino spectrum \(\varepsilon _{\nu,1 }\) can be written as

   \begin{equation}
     \begin{aligned}
    \varepsilon _{\nu,1 } =  \varepsilon _{\nu,1 }^{0}\min \left ( 1\text{, } \left ( \frac{\tau _{p\gamma } ^{p}}{5}  \right ) ^{\frac{1}{1-\beta}} \right ) . 
     \end{aligned}
 \end{equation}
The high-energy break of neutrino spectrum \(\varepsilon _{\nu,2 }\) is caused by the synchrotron cooling of charged pions, which can be calculated as
   \begin{equation}
     \begin{aligned}
       \varepsilon _{\nu,2 } =&\text{ }1.17\times 10^{8}\text{GeV}L_{\rm GRB,52}^{-\frac{1}{2} } \\&\varepsilon _{e,-1 }^{\frac{1}{2}}\varepsilon _{B,-1 }^{-\frac{1}{2}}R_{13}\left ( \frac{\Gamma }{300} \right )^{2} \left ( 1+z \right ) ^{-1}.  
     \end{aligned}
 \end{equation}
where \(\varepsilon _{e}\) and \(\varepsilon _{B}\) are the fraction of jet dissipated energy going into the electrons and the random magnetic fields, respectively.\footnote{In our calculations, we assume \(\varepsilon _{e} = \varepsilon _{B}\).} The specific neutrino number at the low-energy break \(n _{\nu,1 }\) can be calculated as
   \begin{equation}
     \begin{aligned}
   n _{\nu,1 } = \frac{1}{8}\frac{f_{p} f_{\pi }\left ( \varepsilon _{p } \text{/} \varepsilon _{e }\right )E_{\rm GRB}}{\ln_{}{\left ( \varepsilon _{\nu,2 } \text{/}\varepsilon _{\nu,1 } \right )  } }\varepsilon _{\nu,1 }^{-2}, 
     \end{aligned}
 \end{equation}
where \(E_{\rm GRB}\) is the isotropic energy of GRB prompt emission, \(\varepsilon _{p}\) is the fraction of jet dissipated energy that goes into the protons, and \(f_{p}\) is the fraction of accelerated protons that can efficiently participate in \(p\gamma \) interaction. The coefficient $\frac{1}{8}$ comes from the production of $\frac{1}{2}$ and $\frac{1}{4}$, and the $\frac{1}{2}$ is from roughly half of \(p\gamma \) interaction which goes to the $\pi^{+}$ channel, $\frac{1}{4}$ is caused by the $\pi^{+}$ decaying into four different leptons on average. For \(p = 2\), \(f_{p}\) can be expressed as
    \begin{equation}
     \begin{aligned}
     f_{p} = \frac{\ln_{}{\left ( \varepsilon _{\nu,2 } \text{/}\varepsilon _{\nu,1 } \right )}}{\ln_{}{\left ( E _{p,\text{max} } \text{/}E _{p,\text{min} } \right )}},
     \end{aligned}
 \end{equation}
where \(E _{p,\text{max}}\) and \(E _{p,\text{min}}\) are the maximum and minimum energy of accelerated protons in the observer frame, respectively. One can estimate the upper limit of \(E _{p,\text{max}}\) when the dynamical timescale \(t_{\rm dyn}^{'} = R\text{/} \left (\Gamma c  \right ) \) is equal to the accelerating timescale \(t_{\rm acc}^{'} = E_{p}^{'} /(eB^{'}c) \) . Here, \(B^{'} =\left [ \frac{L_{\rm GRB}\left ( \varepsilon _{B } \text{/} \varepsilon _{e }\right )}{2R^{2}\Gamma^{2}c }  \right ] ^{1/2}\) is the strength of the random magnetic field. The lower limit of \(E _{p,\text{min}}\) can also be expressed as \(E _{p,\text{min}}> \Gamma m_{p}c^{2}  \), where \(m_{p}\) is the mass of the proton. Based on the statement before the Equation(\ref{4}), the pion production efficiency \(f_{\pi}\) can be calculated as
    \begin{equation}
     \begin{aligned}
    f_{\pi} =   \begin{cases}
    1-\left ( 1-\left \langle f_{p\to \gamma}  \right \rangle  \right ) ^{\tau _{p\gamma } ^{p}}  , & \tau _{p\gamma }^{p}< 3\\
    1-\left ( 1-\left \langle f_{p\to \gamma}  \right \rangle  \right ) ^{\tau _{p\gamma } ^{p^{2} }}  ,   & \tau _{p\gamma }^{p}\ge  3,
    \end{cases}
     \end{aligned}
 \end{equation}
where \(\left \langle f_{p\to \gamma}\right \rangle = 0.2\) is the fraction of energy transferred from protons to pions. The observed neutrino fluence can be calculated as
    \begin{equation}
     \begin{aligned}
   E_{\nu }^{2} \phi _{\nu}\left ( E_{\nu } \right )  =  \frac{E_{\nu }^{2}n_{\nu } \left ( E_{\nu } \right )}{4\pi d_{L}^{2} } ,
     \end{aligned}
 \end{equation}
where \(d_{L}\) is the luminosity distance.

\section{Constraints on the Physical Parameters Space with Nondetection Neutrinos from GRBs} \label{sec:3}
\begin{figure}[H]
      \centering
      \includegraphics[width=0.495\linewidth]{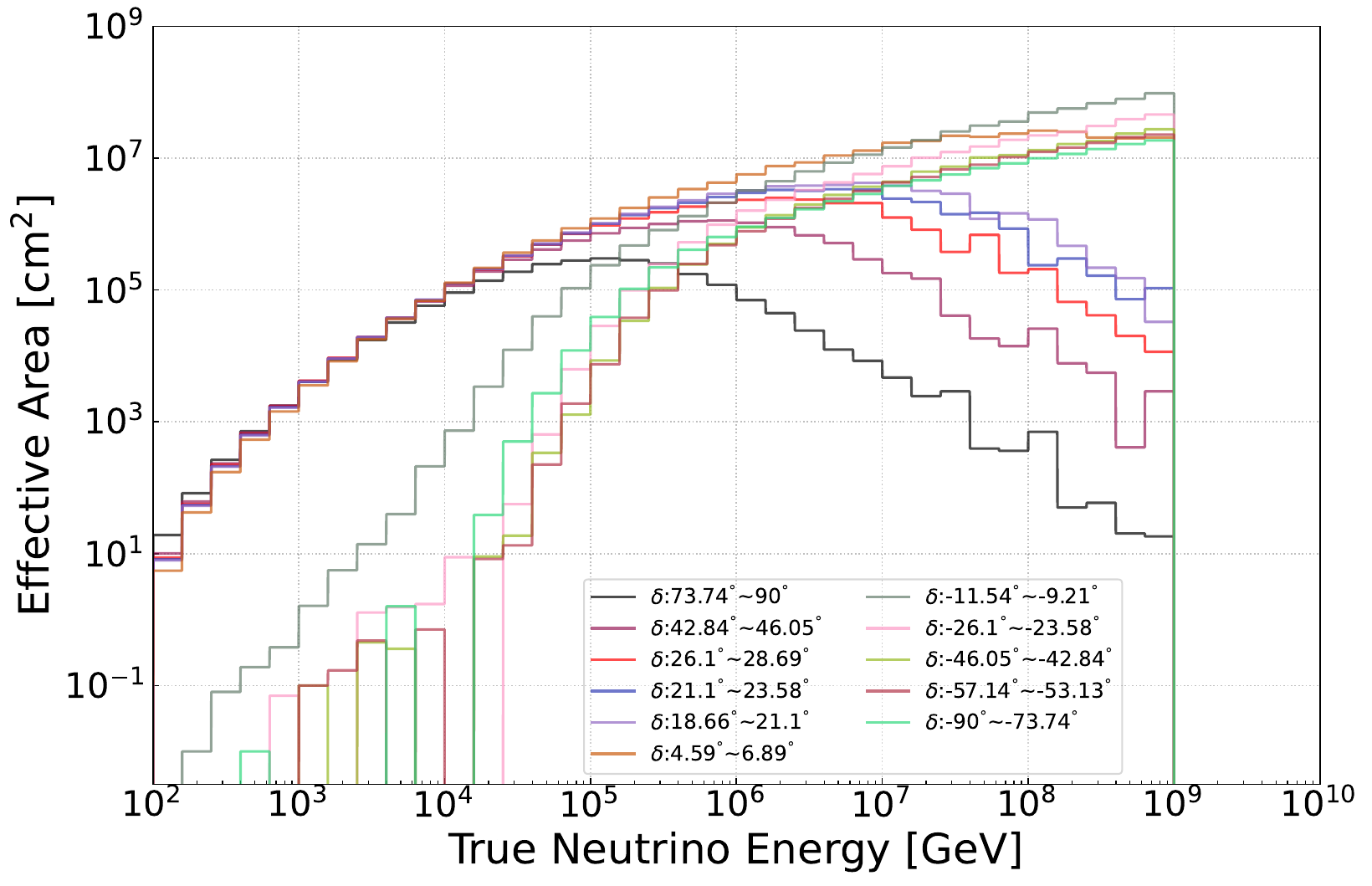}
      \includegraphics[width=0.495\linewidth]{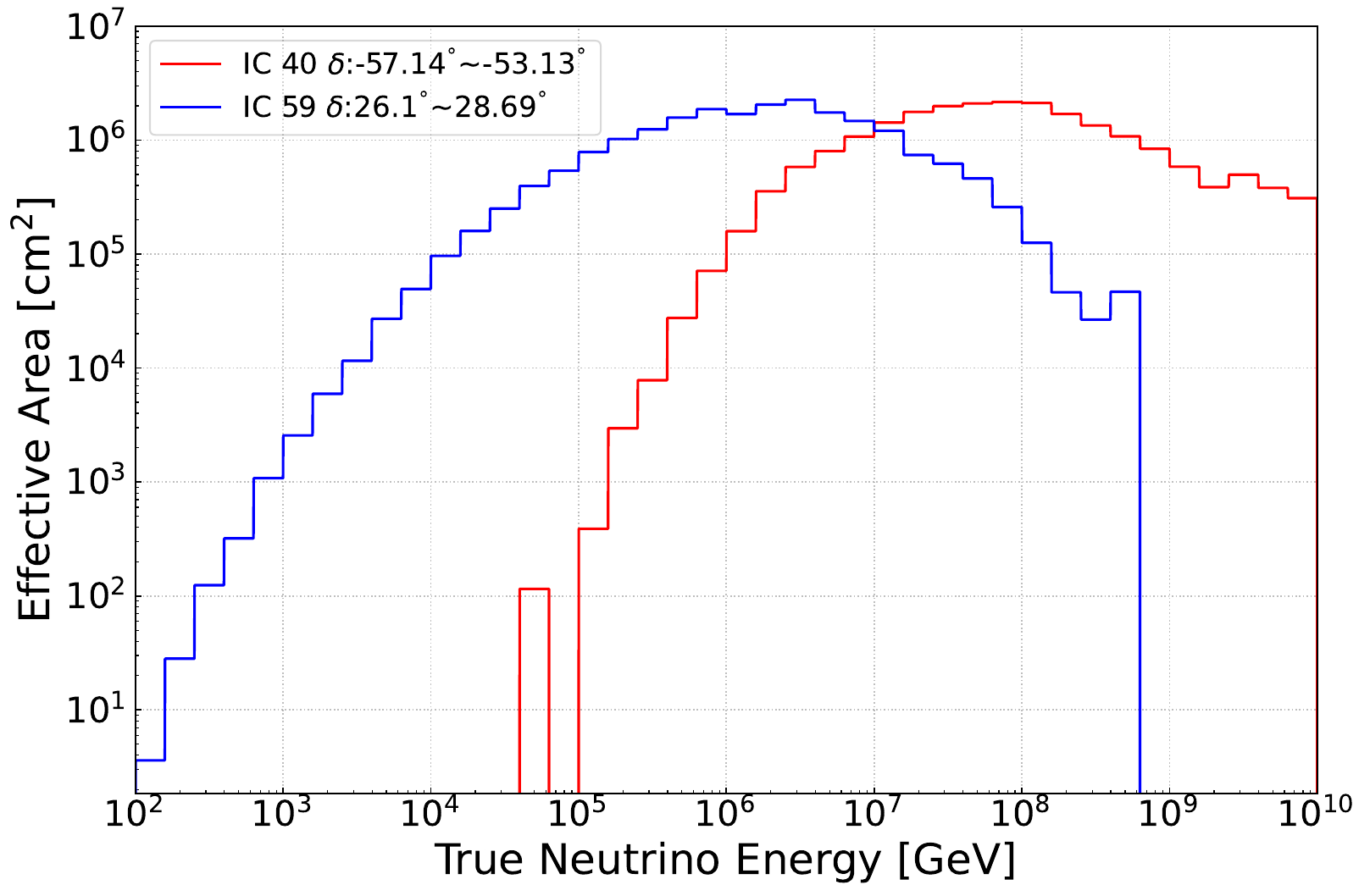}

      \caption{Left: IC86\uppercase\expandafter{\romannumeral+2} effective area as a function of the neutrino energy with different declinations for IceCube. Right: The effective area of other periods used (e.g., IC40 and IC 59) in this paper is teken from \citep{2021arXiv210109836I}}.
      \label{Fig:1}
  \end{figure}

\citet{2024ApJ...964..126A} presented the results of a search for neutrinos from 2268 GRBs by adopting the neutrino spectra with a single power-law model and reported the upper limits from the three most significant GRB–neutrino coincidences. However, the BOAT GRB 221009A invokes the subphotospheric neutron–proton collision model to constrain the limit on baryon loading \citep{2022ApJ...941L..10M,2024ApJ...964..126A}. In this paper, we adopt the effective area of IceCube to calculate the expected number of neutrino signal events and constrain the GRB parameters space. The expected number of neutrino signal events can be calculated by convoluting the neutrino flux and the effective area, one has
    \begin{equation}\label{Eq.N_sig}
     \begin{aligned}
   N_{\rm sig} = \int \phi _{\nu}\left ( E_{\nu } \right )A_{\rm eff} \left ( E_{\nu }, \delta  \right )\mathrm{d}E_{\nu },
     \end{aligned}
 \end{equation}
where \(A_{\rm eff} \) is the effective area of the neutrino detector, and \(\delta\) is the decl. of the source. In our calculations, we adopt the effective area of a 10 year point source \citep{2021arXiv210109836I} to calculate the expected number of neutrino signal events for individual GRBs. In the 10 year point-source dataset, the effective area is changed with different periods, because the different number of strings were operated. The IC86\uppercase\expandafter{\romannumeral+2} effective area and effective area of other periods used in this paper as a function of the neutrino energy for different declinations, which derive from the 10 yr
IceCube dataset \footnote{https://icecube.wisc.edu/data-releases/2021/01/all-sky-point-source-icecube-data-years-2008-2018/}, is shown in Figure \ref {Fig:1}. In this paper, we adopt different effective areas based on the decl. of the burst and the time of the burst happening unless specifically stated \citep[e.g.,][]{ 2012ApJ...752...29H}. The start and end dates of different periods can be found in Table 1 of \cite{2021arXiv210109836I}.

In this section, based on the nonresults of neutrino detection from IceCube Collaboration, we will discuss the constraints of the physical parameters of four peculiar GRBs (080916C, 090902B, 211211A, 230307A), which are from two different types of jet composition (e.g., photosphere origin scenario of GRB 211211A and GRB 090902B, and Poynting-flux-dominated scenario of GRB 230307A and GRB 080916C).

\subsection{Photosphere origin scenario}
\begin{figure}[htbp]
      \centering
      \includegraphics[width=0.8\linewidth]{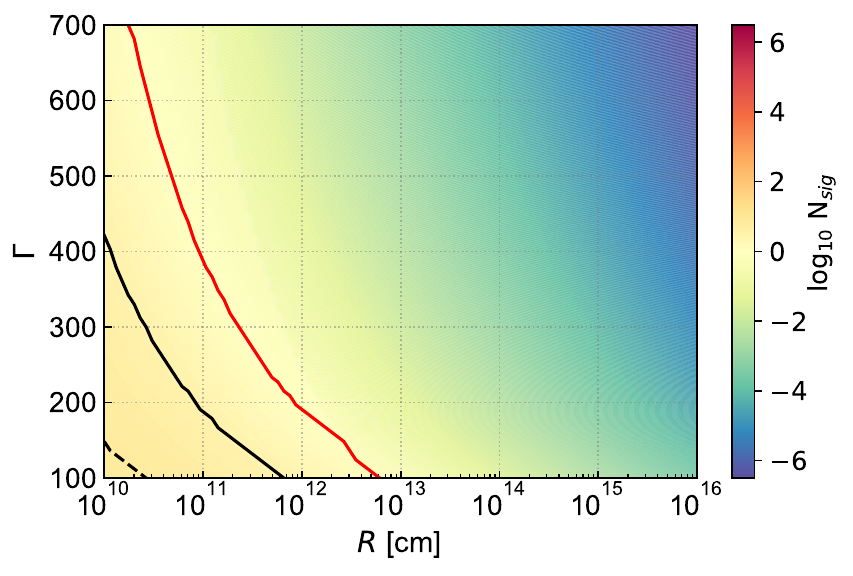}
      \caption{Model-independent constraint on the \(R-\Gamma\) diagram for GRB 211211A. The lines represent \(N_{sig} = 1\) with different GRB parameters, e.g., the solid red line (\(\varepsilon _{p } \text{/} \varepsilon _{e } = 10\) and \(f_{p} = 1\)); the solid black line (\(\varepsilon _{p } \text{/} \varepsilon _{e } = 3\) and \(f_{p} = 1\)); the dotted black line (\(\varepsilon _{p } \text{/} \varepsilon _{e }  = 3\) and \(f_{p} = 0.5\)).} 
      \label{Fig:2}
  \end{figure}

GRB 211211A, a short GRB with extended emission possibly associated with kilonova, is believed to be from the merger of compact stars \citep{2022Natur.612..223R,2022Natur.612..228T,2022Natur.612..232Y,2023ApJ...943..146C,2023NatAs...7...67G}. The thermal component is indeed existent in the prompt emission \citep{2023ApJ...943..146C} but does not detect any neutrino counterparts. The decl. of GRB 211211A is \(\delta \approx 27.9^{\circ}\) which can be used to estimate the effective areas based on Figure \ref {Fig:1}. By adopting the spectral parameters of GRB 211211A, \(\alpha = 1.2\), \(\beta = 2.05\), break energy  \(\varepsilon _{\gamma,b } = 0.499\text { MeV} \), \(E_{\rm GRB} = 7.6\times10^{51}\text { erg}\), isotropic luminosity \(L_{\rm GRB} = 1.1\times10^{50}\text { erg/s}\) and redshift \(z = 0.076\) \citep{2022Natur.612..232Y}, together with the formulas in Section\ref{sec:2} and Equation(\ref{Eq.N_sig}), one can present the model-independent constraints in the \(R-\Gamma\) diagram of GRB 211211A with different \(\varepsilon _{p } \text{/} \varepsilon _{e }\) and \(f_{p}\) (see Figure \ref {Fig:2}). 

\begin{figure}[htbp]
      \centering
      \includegraphics[width=0.495\linewidth]{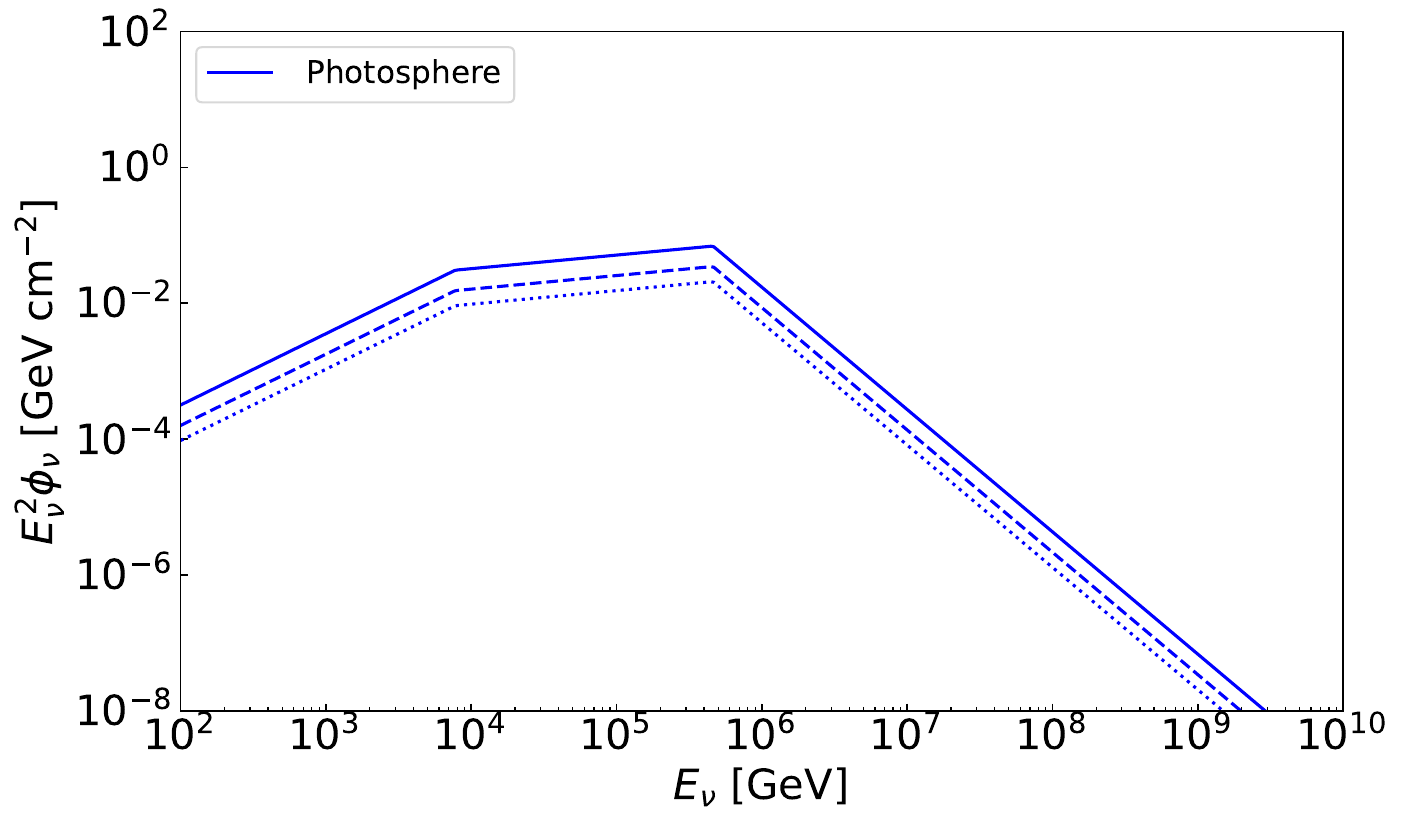}
      \includegraphics[width=0.495\linewidth]{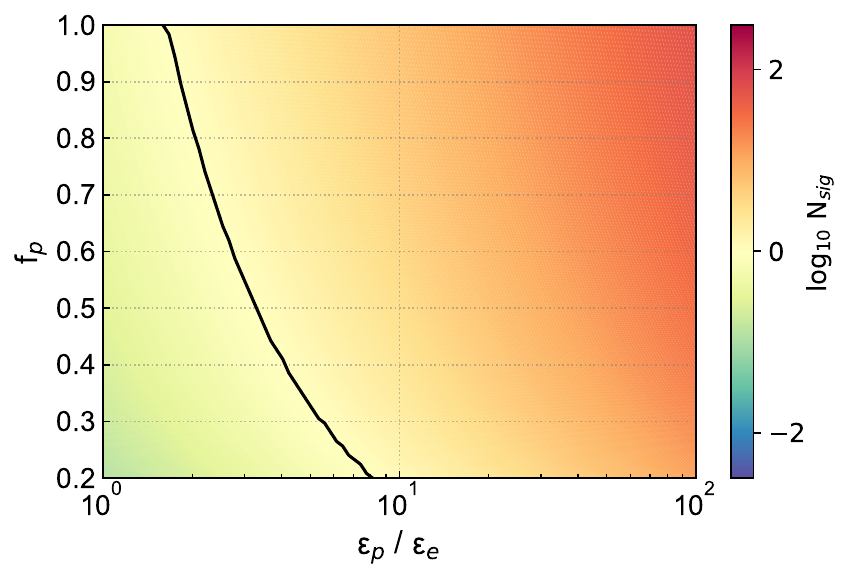}
      \caption{Left: predicted neutrino spectra of GRB 211211A for the dissipative photosphere model with \(R_{\text{ph}}=10^{10}\text{cm}\), \(\Gamma = 200\), and \(\varepsilon _{p } \text{/} \varepsilon _{e } = 3\). Three lines are corresponding to different values of \(f_{p}\), e.g., solid line (\(f_{p}=1.0\)), dashed line (\(f_{p}=0.5\)), and dotted line (\(f_{p}=0.3\)).
      Right: constraint on \(\varepsilon _{p } \text{/} \varepsilon _{e }\) of GRB 211211A within dissipative photosphere model for \(R_{\text{ph}}=10^{10}\text{cm}\) and \(\Gamma = 200\).}
      \label{Fig:3}
  \end{figure}

\cite{2023ApJ...943..146C} calculated the photosphere radius \(R_{\text{ph}}=10^{10}\text{cm}\) and the bulk Lorentz factor \(\Gamma = 200\) of GRB 211211A based on the thermal emission. By adopting the above parameters of GRB 211211A, we find that only the dissipative photosphere model can be well constrained. Figure \ref {Fig:3} shows the neutrino spectra and constraints on \(\varepsilon _{p } \text{/} \varepsilon _{e }\) for GRB 211211A in the dissipative photosphere model, and it requires \(\varepsilon _{p } \text{/} \varepsilon _{e }<8\) for \(f_{p}>0.2\). From the theoretical point of view, the required \(10 < \varepsilon _{p } \text{/} \varepsilon _{e } \le 100\) which depended on the energy generation rate of both UHECR and GRBs in the GRB–UHECR hypothesis, are studied by previous works \citep[e.g.,][]{2008PhRvD..78b3005M,2015APh....62...66B,2023ApJ...950...28R}. The value of \(\varepsilon _{p } \text{/} \varepsilon _{e }\) in the dissipative photosphere model may be lower than that in the internal shock model, and we cannot rule out the photosphere origin. In general, cosmic rays are difficult to accelerate into ultrahigh energies through subphotosphere dissipation within the scenario of a dissipative photosphere model \citep{2008PhRvD..78j1302M}. Even if, it also requires a larger value of \(\varepsilon _{p } \text{/} \varepsilon _{e }\). So, the UHECR hypothesis for GRB 211211A in the dissipative photosphere model could be ruled out.

Another case with thermal-dominated outflow is GRB 090902B \citep{2010ApJ...709L.172R}. The decl. of GRB 090902B is \(\delta \approx 27.9^{\circ}\), which is similar to GRB 211211A. The time-integrated spectrum of GRB 090902B can be well fitted by blackbody(BB) and power-law (PL) models. By assuming that the PL component is a low-energy region in the high-energy cutoff model. Thus, we adopt the index of the PL component as the low-energy index of the Band function to calculate neutrino emission.

\begin{figure}[H]
      \centering
      \includegraphics[width=0.495\linewidth]{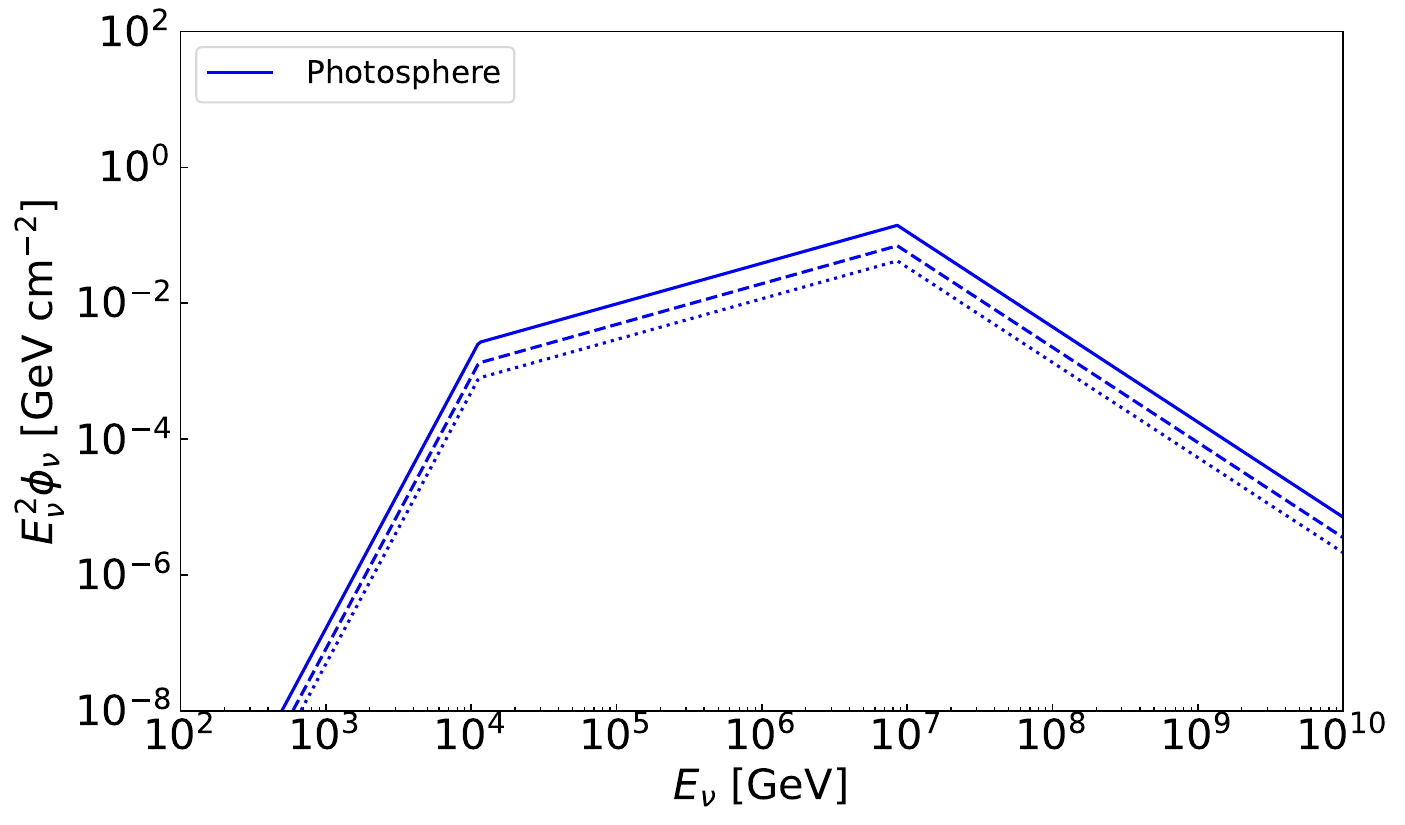}
      \includegraphics[width=0.495\linewidth]{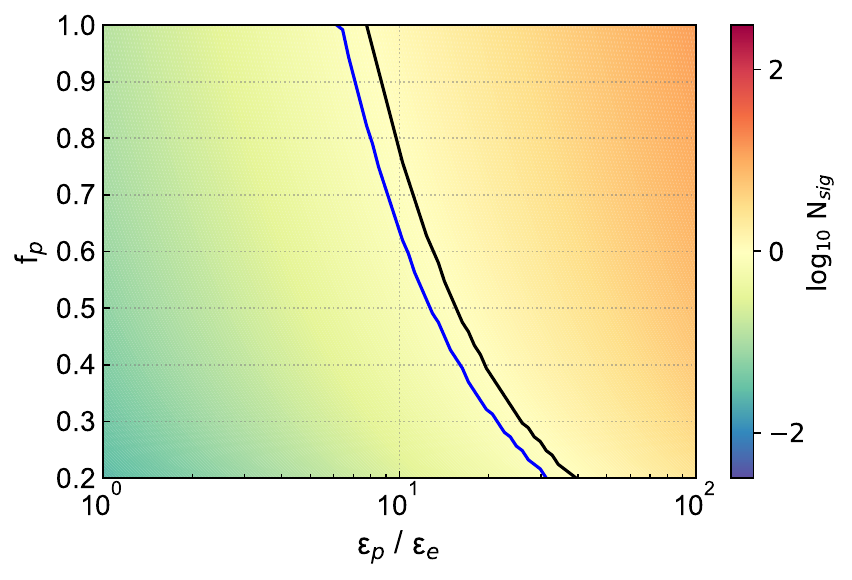}
      \caption{Left: predicted neutrino spectra of GRB 090902B for the dissipative photosphere model with \(R_{\text{ph}}=1.1\times10^{12}\text{cm}\), \(\Gamma = 580\), and \(\varepsilon _{p } \text{/} \varepsilon _{e } = 3\). Three lines are corresponding to different values of \(f_{p}\), e.g., solid line (\(f_{p}=1.0\)), dashed line (\(f_{p}=0.5\)), and dotted line (\(f_{p}=0.3\)).
      Right: constraint on \(\varepsilon _{p } \text{/} \varepsilon _{e }\) of GRB 090902B within dissipative photosphere model for \(R_{\text{ph}}=1.1\times10^{12}\text{cm}\) and \(\Gamma = 580\). The solid black and blue lines are \(N_{\rm sig}=1\) at \(\delta\approx 27.9^{\circ}\) by adopting the effective area of IC59 and IC86\uppercase\expandafter{\romannumeral+2}, respectively.}
      \label{Fig:4}
  \end{figure}

The spectral parameters of GRB 090902B prompt emission can be presented as follows, \(\alpha = 1.6\), \(E_{\rm GRB} = 3.63\times10^{54}\text { erg}\), \(L_{\rm GRB} = 4\times10^{52}\text { erg/s}\), and redshift \(z = 1.822\) \citep{2010ApJ...709L.172R}. Since we cannot obtain the \(\beta\) and \(\varepsilon _{\gamma,b }\) from the observations of GRB 090902B, and we note that the adopted varying of \(\beta\) (e.g., from 3 to 5) cannot significantly affect the \(N_{\rm sig}\). However, increasing the value of \(\varepsilon _{\gamma,b }\) can result in a decreasing value of \(N_{\rm sig}\). So, we adopt \(\beta = 5\) and the lower limit of \(\varepsilon _{\gamma,b } = 30\text { MeV} \) to do the calculations. Moreover, the \(R_{\text{ph}}=1.1\times10^{12}\text{cm}\) and \(\Gamma = 580\) are adopted based on the results in \cite{2010ApJ...709L.172R}, and then calculate the neutrino emission from photosphere and \(N_{\rm sig}\) by using IceCube 59-string effective area (IC59) and IC86\uppercase\expandafter{\romannumeral+2}. The neutrino spectra and constraints on \(\varepsilon _{p } \text{/} \varepsilon _{e }\) for GRB 090902B in the dissipative photosphere model are shown in Figure \ref{Fig:4}. It is found that \(\varepsilon _{p } \text{/} \varepsilon _{e }<40\) for \(f_{p}>0.2\) by adopting IC59.

\subsection{Poynting-flux-dominated scenario}
GRB 230307A, also classified as short GRB with extended emission, is claimed to be from the merger of compact stars \citep{2023arXiv230705689S,2024ApJ...962L..27D}. The lack of detected thermal emission and mini-structure of prompt emission suggest that the outflow is Poynting-flux-dominated \citep{2023arXiv231007205Y,2024MNRAS.529L..67D}. 

The IceCube Collaboration has performed dedicated analyses to search for high-energy neutrino events from GRB 230307A \citep{2023GCN.33430....1I}. By assuming a single power-law \(E_{\nu }^{-2}\) spectral within the energy range from 90 TeV to 20 PeV, the IceCube Collaboration reports a time-integrated muon-neutrino flux upper limit of 1.0 GeV cm\(^{-2}\) at 90\% confidence level \citep{2023GCN.33430....1I}. It is worth
noting that the IceCube upper limit is available only for GRB 230307A among the four selected GRBs. From the theoretical point of view, for GRB 230307A, the ICMART model predicted that the neutrino energy may be larger than 20 PeV, and the effective area of IceCube in the higher energy region (above 20 PeV) is larger than that of  the range from 90 TeV to 20 PeV.

\begin{figure}[htbp]
      \centering
      \includegraphics[width=0.495\linewidth]{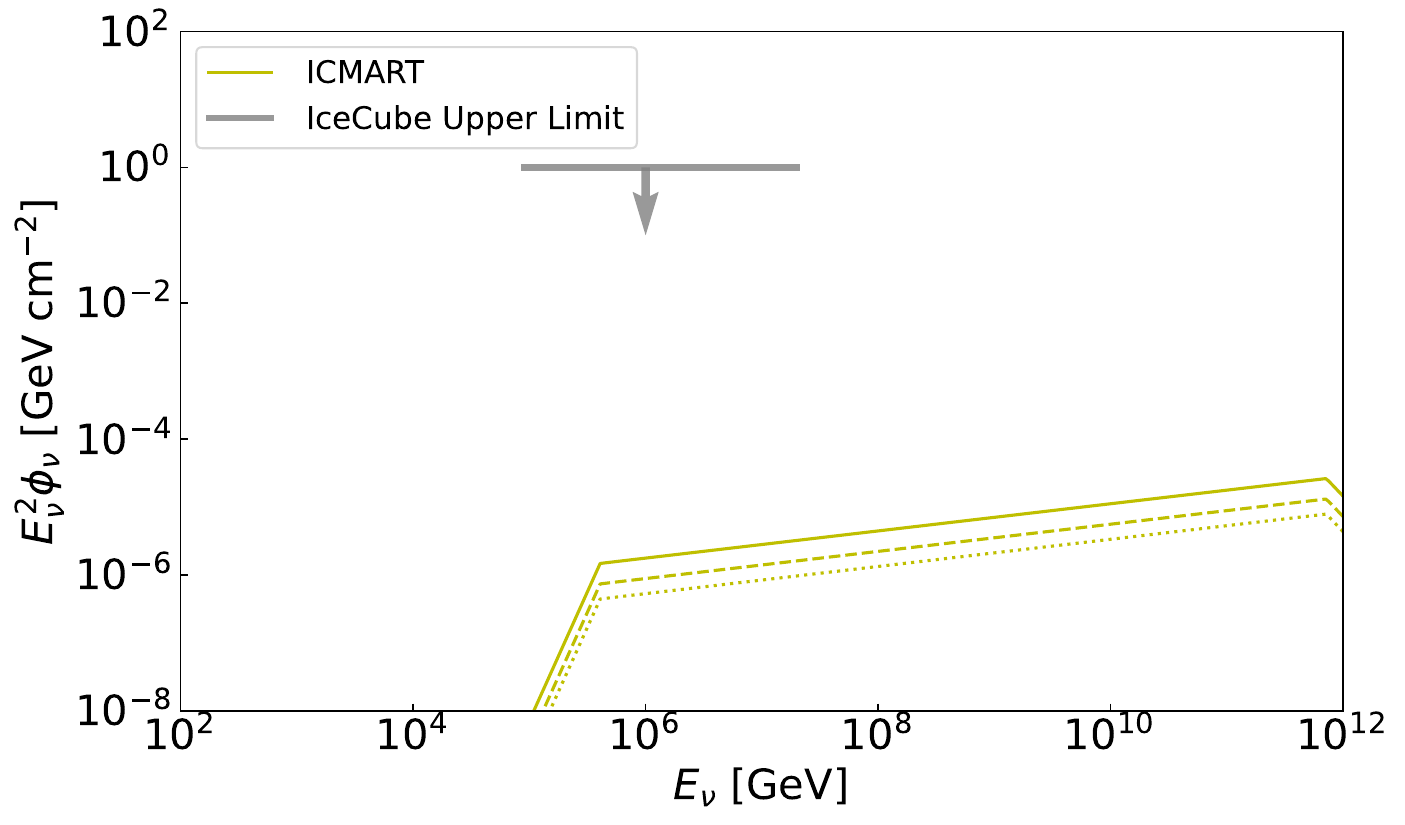}
      \includegraphics[width=0.495\linewidth]{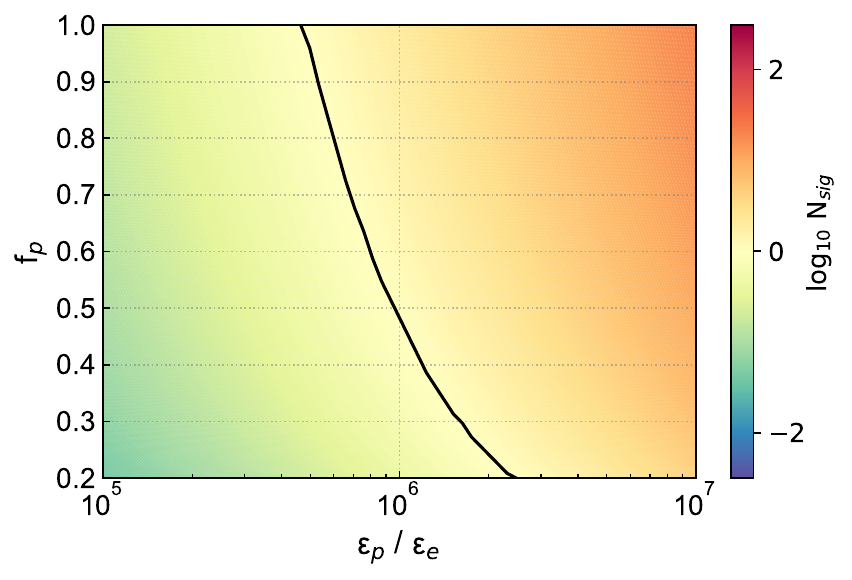}
      \caption{Left: predicted neutrino spectra of GRB 230307A for the ICMART model with \(R_{\text{ICMART}}=10^{16}\text{cm}\), \(\Gamma = 300\), and \(\varepsilon _{p } \text{/} \varepsilon _{e } = 3\). Three lines are corresponding to different values of \(f_{p}\), e.g., solid line (\(f_{p}=1.0\)), dashed line (\(f_{p}=0.5\)), and dotted line (\(f_{p}=0.3\)). The gray line with arrows is the upper limit of time-integrated muon-neutrino flux for IceCube \citep{2023GCN.33430....1I}.
      Right: constraint on \(\varepsilon _{p } \text{/} \varepsilon _{e }\) of GRB 230307A for the ICMART model with \(R_{\text{ICMART}} = 10^{16}\text{cm}\) and \(\Gamma = 300\).} 
      \label{Fig:5}
  \end{figure}

Due to the presence of pulse pile-up in GBM, the spectra of the prompt emission of GRB 230307A after 2.5s will be distorted \citep{2023GCN.33551....1D}, therefore, we only adopt the first 2.5s data of GBM to do the spectral analysis, and determine the high-energy photon spectral index \(\beta\).
Based on results from \cite{2023arXiv230705689S}, we adopt the parameters of GRB 230307A to do the calculations, e.g., \(\alpha = 1.2\), \(\beta = 4.8\), \(\varepsilon _{\gamma,b } = 1.563\text { MeV} \), \(E_{\rm GRB} = 3.08\times10^{52}\text { erg}\), \(L_{\rm GRB} = 2.35\times10^{50}\text { erg/s}\), and redshift \(z = 0.065\). Moreover, based on the method in \citet{}{2009ApJ...700L..65Z} and the result in \citet{2024MNRAS.529L..67D}, we adopt the \(R_{\text{ICMART}}=10^{16}\text{cm}\) and bulk Lorentz factor \(\Gamma = 300\). The neutrino spectra and constraints on \(\varepsilon _{p } \text{/} \varepsilon _{e }\) for GRB 230307A are shown in Figure \ref{Fig:5}. It is found that the predicted neutrino fluence of GRB 230307A in the ICMART model is much lower than that of the upper limit of IceCube. The upper limit of \(\varepsilon _{p } \text{/} \varepsilon _{e }\) is much larger than that of a typical value. Therefore, the nondetected result of the neutrino counterpart of GRB 230307A cannot effectively constrain the ICMART model.

Another case of Poynting-flux-dominated outflow is GRB 080916C which lacks the thermal emission \citep{2009Sci...323.1688A,2009ApJ...700L..65Z}. The IceCube was under construction when GRB 080916 was detected at that time. The IceCube 40-string observation still gives us an opportunity to constrain the ICMART model. Here, we adopt the parameters of GRB 080916C as following, \(\alpha = 1.07\), \(\beta = 2.05\), \(\varepsilon _{\gamma,b } = 0.667\text { MeV} \),  \(E_{\rm GRB} = 8.8\times10^{54}\text { erg}\), \(L_{\rm GRB} = 7\times10^{52}\text { erg/s}\), and redshift \(z = 4.35\) \citep{2009Sci...323.1688A}. The values of lower limits of \(R\) and \(\Gamma \) are taken from \cite{2009ApJ...700L..65Z} to calculate the neutrino emission from the ICMART model, such as \(R_{\text{ICMART}} = 10^{16}\text{cm}\) and \(\Gamma = 500\).

\begin{figure}[htbp]
      \centering
      \includegraphics[width=0.495\linewidth]{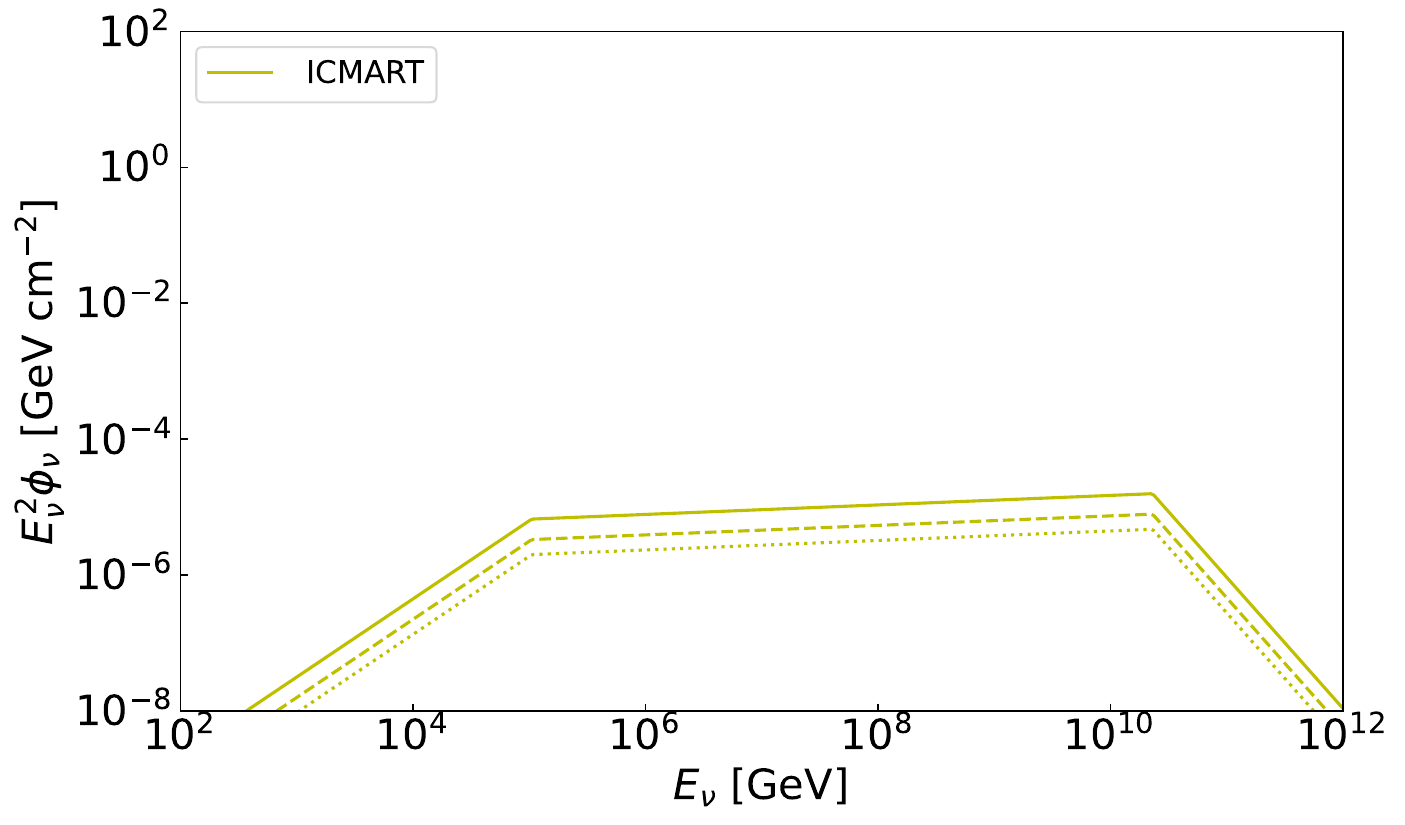}
      \includegraphics[width=0.495\linewidth]{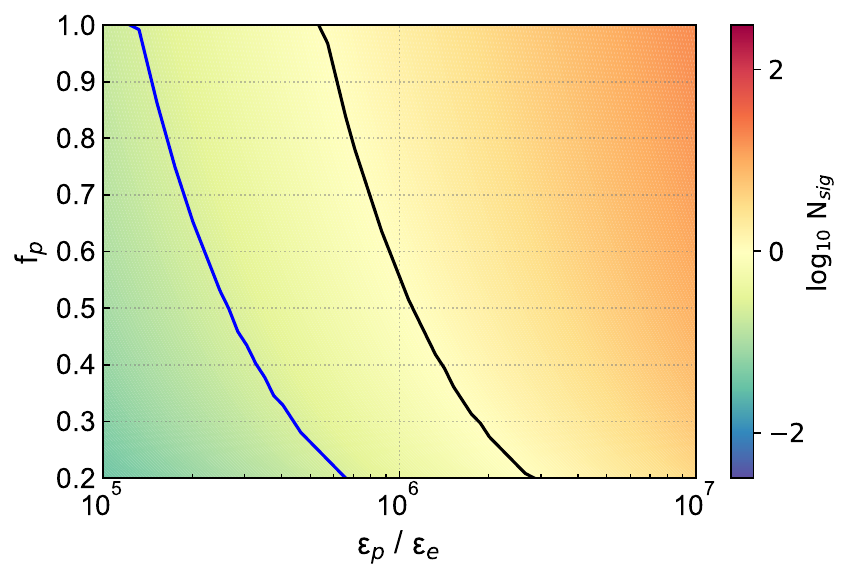}
      \caption{Left: predicted neutrino spectra of GRB 080916C for the ICMART model with \(R_{\text{ICMART}}=10^{16}\text{cm}\), \(\Gamma = 500\), and \(\varepsilon _{p } \text{/} \varepsilon _{e } = 3\). Three lines are corresponding to different values of \(f_{p}\), e.g., solid line (\(f_{p}=1.0\)), dashed line (\(f_{p}=0.5\)), and dotted line (\(f_{p}=0.3\)).
      Right: constraint on \(\varepsilon _{p } \text{/} \varepsilon _{e }\) of GRB 080916C for the ICMART model with \(R_{\text{ICMART}} = 10^{16}\text{cm}\) and \(\Gamma = 500\). The solid black and blue lines are \(N_{\rm sig}=1\) at \(\delta\approx -56.6^{\circ}\) by adopting the effective areas of IC40 and IC86\uppercase\expandafter{\romannumeral+2}, respectively.}
      \label{Fig:6}
  \end{figure}

Figure \ref{Fig:6} shows the neutrino spectra and constraints on \(\varepsilon _{p } \text{/} \varepsilon _{e }\) for GRB 080916C in the ICMART model. It is found that the upper limits of \(\varepsilon _{p } \text{/} \varepsilon _{e }\) are much higher than that of the typical value, so there are no effective constraints on the ICMART model of GRB 080916C by using both IceCube effective area, (e.g., IC40 and IC86\uppercase\expandafter{\romannumeral+2}). 

Recently, on the other hand, the IceCube collaboration used the subphotosopheric model to predict the sub-TeV neutrino emission from proton–neutron collision to constraint the baryon loading on GRB 221009A \citep{2024ApJ...964..126A}. The definition of the baryon loading in their work is the ratio of the equivalent isotropic energies in baryons to gamma rays \citep[$\xi_N$, also called the nucleon loading factor;][]{2022ApJ...941L..10M}, and it is similar to the ratio of the total dissipated energy that goes into the protons and electrons \citep[$\varepsilon _{p } \text{/} \varepsilon _{e }$, also called cosmic ray loading factor;][]{2022ApJ...941L..10M}. The equivalent isotropic energies of proton $E_{\rm proton}$ can be calculated by $E_{\rm GRB}\times\left ( \varepsilon _{p } \text{/} \varepsilon _{e }\right )$. For comparison, we adopt our method to calculate the upper limit of \(\varepsilon _{p } \text{/} \varepsilon _{e }\) for GRB 221009A within the internal shock model
\footnote{We adopt the parameters of GRB 221009A as following, \(\delta \approx 19.8^{\circ}\), \(\alpha = 0.97\), \(\beta = 2.34\), \(\varepsilon _{\gamma,b } = 1\text { MeV} \),  \(E_{\rm GRB} = 2\times10^{54}\text { erg}\), \(L_{\rm GRB} = 1.9\times10^{52}\text { erg/s}\), and redshift \(z = 0.151\). The internal shock radius is $R_{\rm IS} = 5.4 \times 10^{14}~{\rm cm}~ (\Gamma/300)^2 (\delta t_{\rm min} / 0.1~{\rm s}) (1+z)^{-1}$ , where $\delta t_{\rm min} = 0.01\rm s$ is the minimum variability time scale for GRB 221009A \citep{2023ApJ...944..115A}.}
, and find \(\varepsilon _{p } \text{/} \varepsilon _{e }<\) 2.02 (465) for a bulk Lorentz factor \(\Gamma = \) 300 (800) at the 90\% confidence level. The results from the
IceCube Collaboration are $\xi_N<$ 3.85 (2.13) for a bulk Lorentz factor \(\Gamma = \) 300 (800) at 90\% confidence level. Although the isotropic energy we adopted in our calculation is lower than that of IceCube Collaboration work, the comparison is still available. By comparing to the subphotosopheric model, our method is in favor of constraint $\varepsilon _{p } \text{/} \varepsilon _{e }$ at lower values of the bulk Lorentz factor but is not good enough at higher values of the bulk Lorentz factor. On the contrary, the constraint from the subphotosopheric model is better than that of our result when the Lorentz factor is increased. Thus, for those two different physical mechanism methods, the constraints of baryon loading from the subphotosopheric model can place a complement to the constraints from photomeson interaction at high values of the Lorentz factor.

\section{Detection Prospects of High-energy Neutrinos from GRBs}\label{sec:4}
In this section, we investigate the detection prospects of high-energy neutrinos from GRBs based on the effective areas of IceCube and IceCube-Gen2. The effective area of IceCube-Gen2 is about 5 times that of IceCube \citep{2021JPhG...48f0501A}. Therefore, we scale IC86\uppercase\expandafter{\romannumeral+2} to estimate the effective area of IceCube-Gen2. Since the effective area strongly depends on the decl. of sources, we will estimate the expected number of neutrino signal events and their detection probability from GRBs. The effective area of IC86\uppercase\expandafter{\romannumeral+2} for different declinations is shown in Figure \ref{Fig:1}. It is found that the effective area of IceCube in the equatorial plane is larger than that of other decl. of sources. Moreover, the effective area of the southern hemisphere in the low-energy range is much lower than that of the northern hemisphere and equatorial plane.

The probability of detecting more than one neutrino can be expressed as \citep{2017ApJ...848L...4K,2023ApJ...950..190M}
    \begin{equation}\label{Eq.P_sig}
     \begin{aligned}
        P({N_{\rm sig} \geq 1}) = 1- e^{-{N_{\rm sig}}}.
     \end{aligned}
    \end{equation}
Here, we only discuss two types of GRB jet compositions, e.g., photosphere with matter-dominated (i.e., GRB 211211A-like and GRB 090902B-like events) and ICMART with Poynting-flux-dominated (i.e., GRB 230307A-like and GRB 080916C-like events). However, the value of \(\varepsilon _{p } \text{/} \varepsilon _{e }\) is difficult to obtain from the observations, we have to assume \(\varepsilon _{p } \text{/} \varepsilon _{e } = 3\) for each GRB in our calculations \citep[e.g.,][]{2023ApJ...944..115A}.

\subsection{Photosphere scenario}
\begin{figure}[H]
      \centering
      \includegraphics[width=0.495\linewidth]{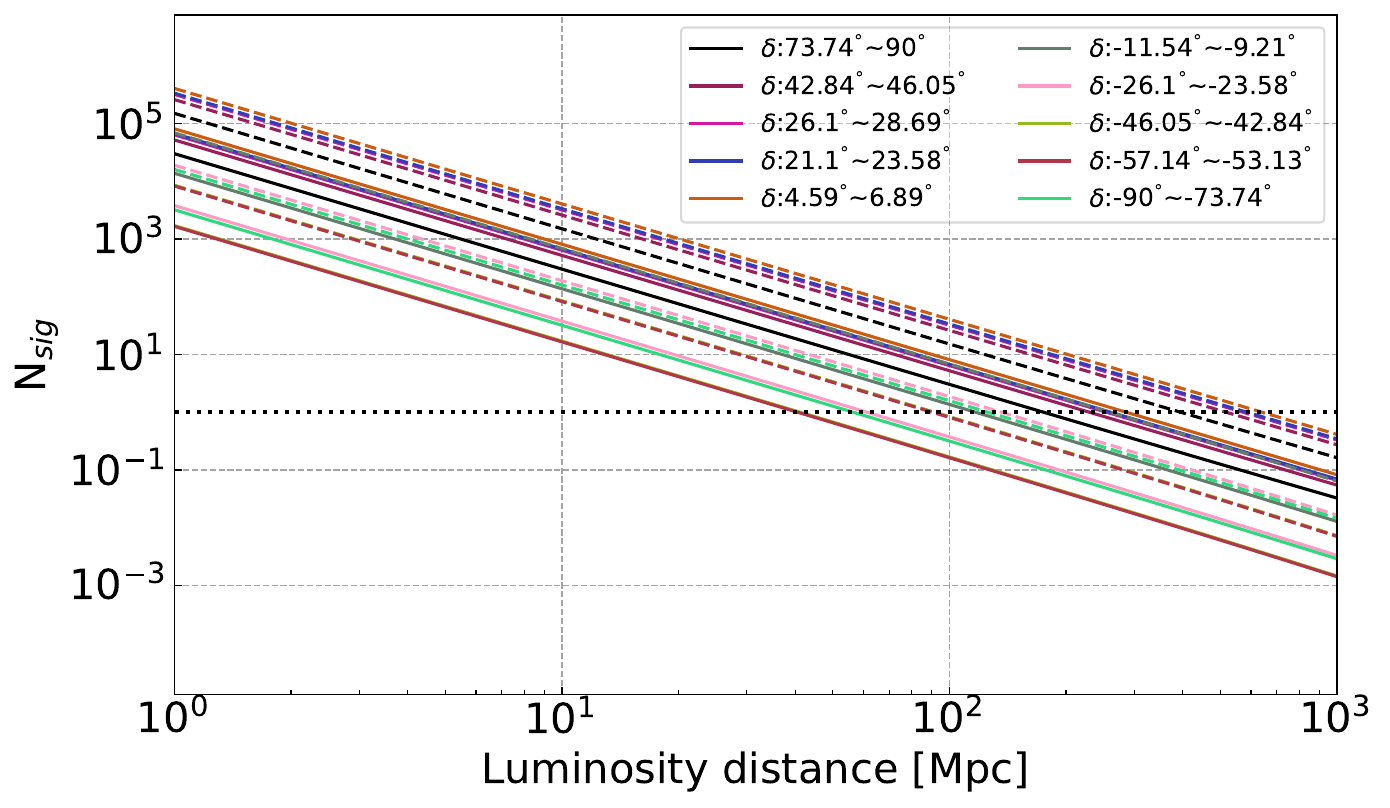}
      \includegraphics[width=0.495\linewidth]{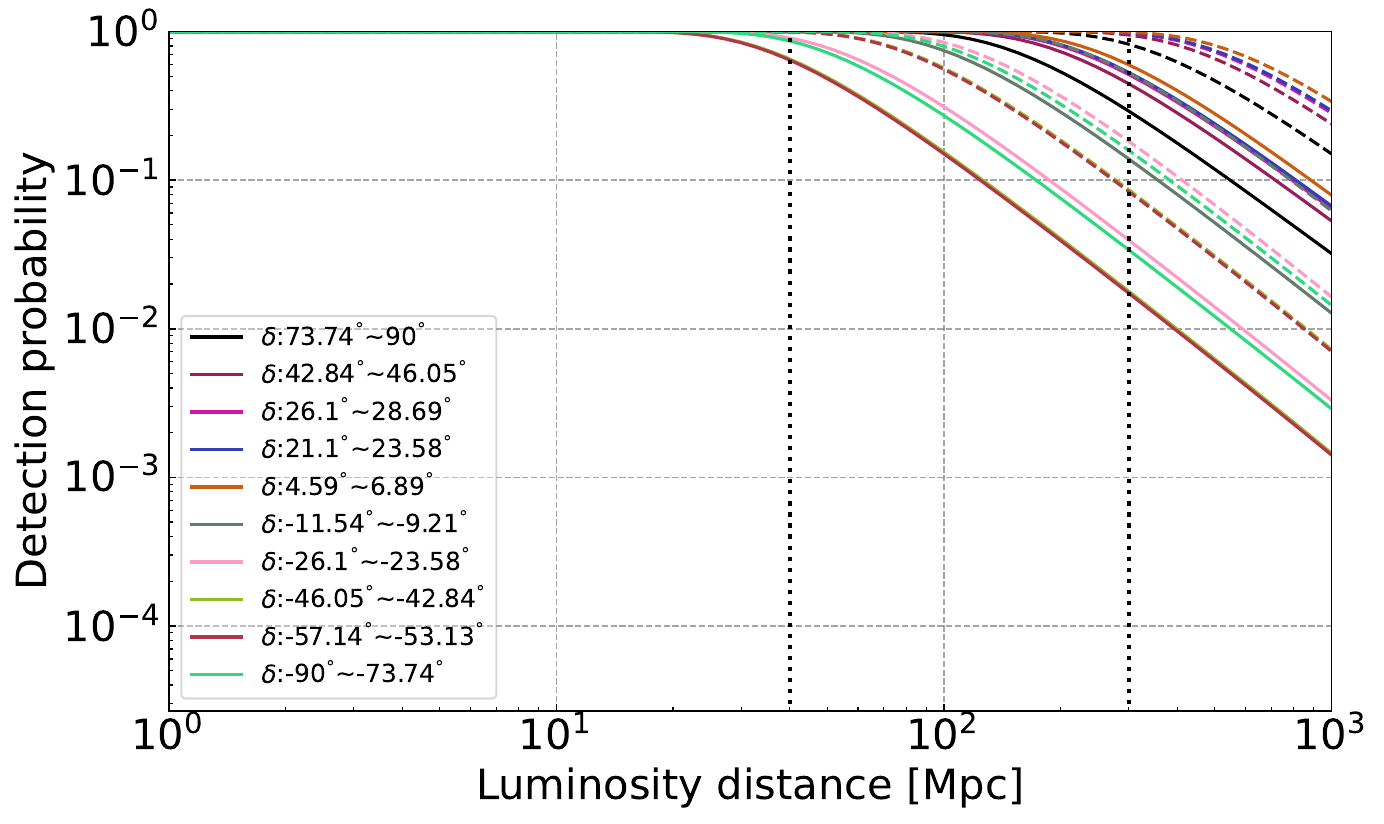}
      \caption{Left: expected number of neutrino signal events for a GRB 211211A-like event as a function of the luminosity distance. The horizontal dotted line is \(N_{\rm sig} = 1\).
      Right: probability of detecting more than one neutrino associated with a GRB 211211A-like event as a function of the luminosity distance. The vertical dotted lines are \(d_{L} = 40\) Mpc and \(d_{L} = 300\) Mpc. The solid and dashed lines are corresponding to adopt the effective area of IceCube and IceCube-Gen2 for different declinations, respectively. \(f_{p} = 0.3\) and \(\varepsilon _{p } \text{/} \varepsilon _{e } = 3\) are adopted.}
      \label{Fig:7}
  \end{figure}

Based on Sections \ref{sec:2} and \ref{sec:3}, and Equations (\ref{Eq.N_sig}) and (\ref{Eq.P_sig}), we investigate the detection prospects of high-energy neutrinos from a GRB 211211A-like event which is thought to be from photosphere emission, and the results are shown in Figure \ref{Fig:7}. The vertical dotted lines are \(d_{L} = 40\) Mpc and \(d_{L} = 300\) Mpc which correspond to the distance of the discovered nearest GRB 170817A and the upper limit of gravitational wave (GW) detectors (e.g., aLIGO) for mergers of the binary neutron star. For an ideal case, it is found that at least one high-energy neutrino associated with a GRB 211211A-like event is possibly detected at \(d_{L} \le 300\) Mpc for IceCube, and \(d_{L} \le 700\) Mpc for IceCube-Gen2. These results indicate that it is possible to detect at least one neutrino coincident with GW from a GRB 211211A-like event in the future if such an event originates from mergers of compact stars with photosphere dissipation.

\begin{figure}[H]
      \centering
      \includegraphics[width=0.495\linewidth]{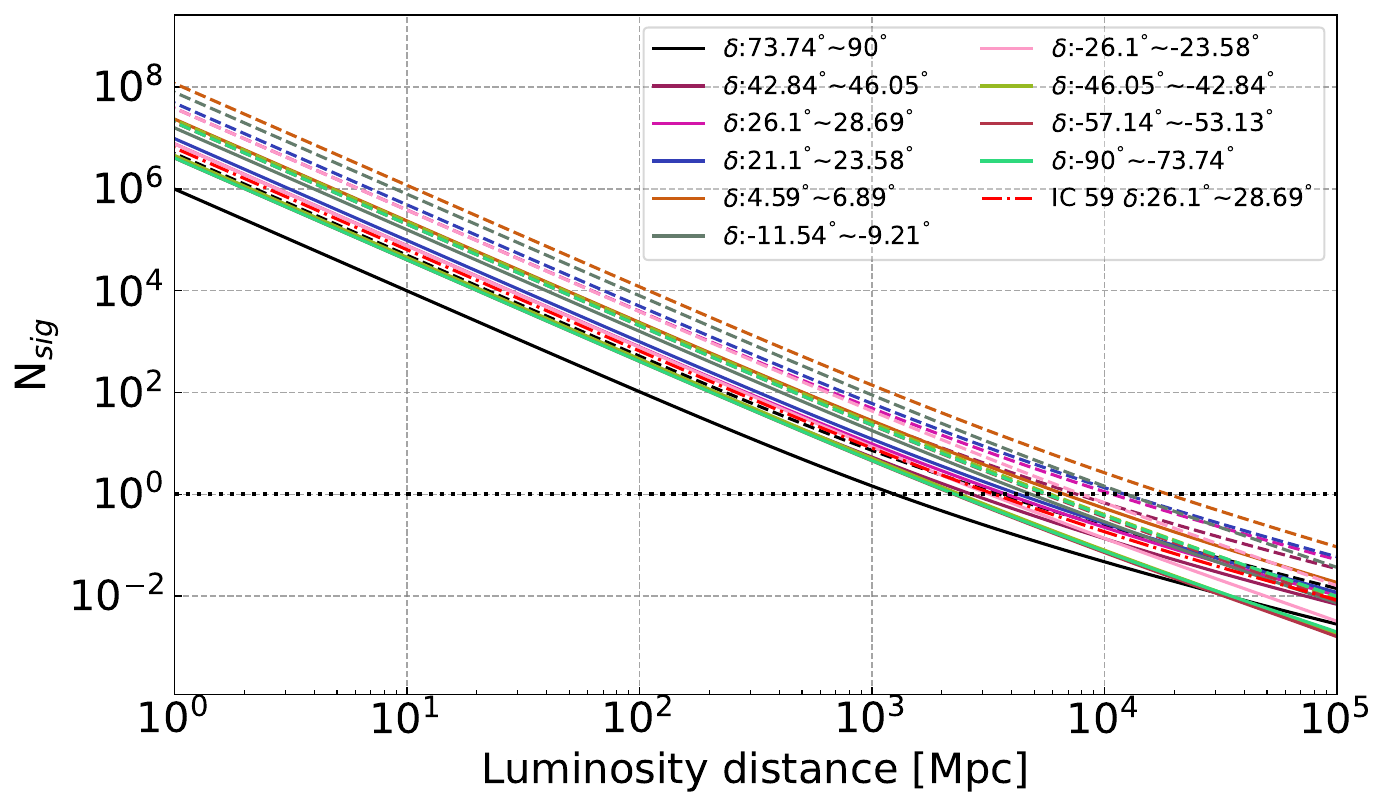}
      \includegraphics[width=0.495\linewidth]{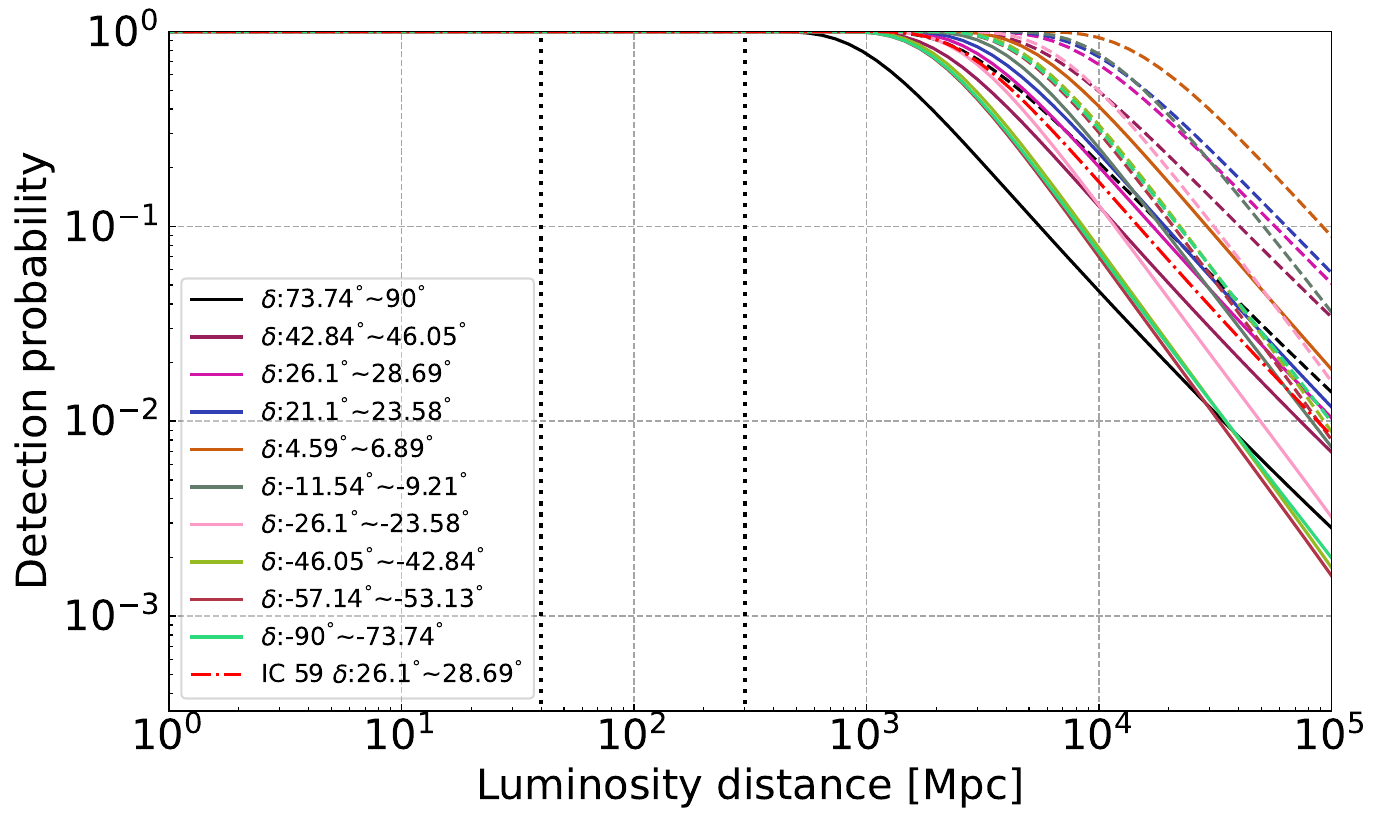}
      \caption{Similar to Figure \ref{Fig:7}, but for a GRB 090902B-like event.}
      \label{Fig:8}
  \end{figure}

Moreover, we also investigate the detection prospects of high-energy neutrinos from another GRB 090902B-like event with photosphere emission, and the results are shown in Figure \ref{Fig:8}. Because of the high isotropic energy of GRB 090902B and the high efficiency of neutrino production near the photosphere, the detection horizon of neutrinos from GRB 090902B-like events can reach several Gpc for IceCube and 20 Gpc for IceCube-Gen2. The redshift of GRB 090902B is 1.822 which is the corresponding luminosity distance \(d_{L} \sim 13\) Gpc. We cannot detect neutrinos from GRB 090902B-like events for IceCube if the proton acceleration is not strong enough near the photosphere. However, it is possible to be detected from GRB 090902B-like events for IceCube-Gen2.

\subsection{Poynting-flux-dominated scenario}
\begin{figure}[H]
      \centering
      \includegraphics[width=0.495\linewidth]{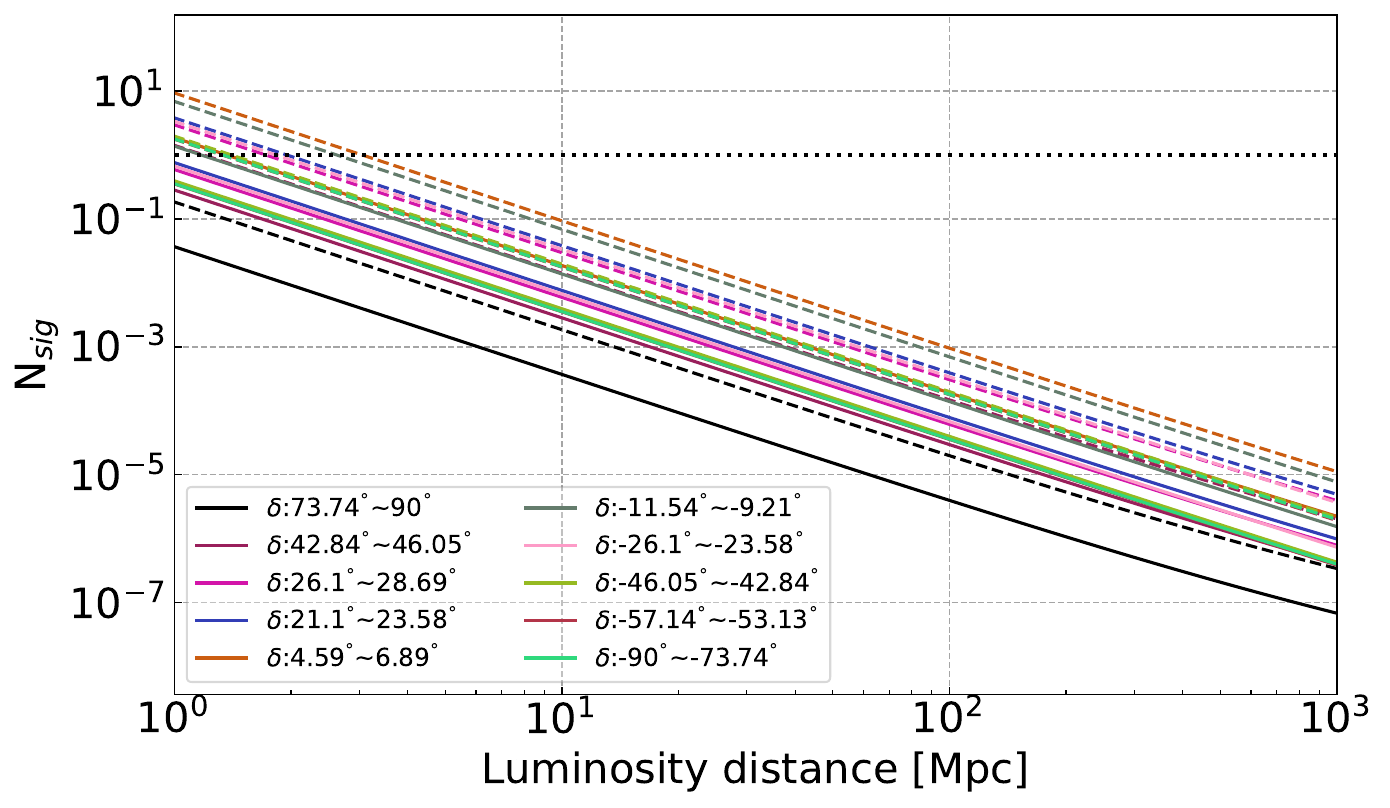}
      \includegraphics[width=0.495\linewidth]{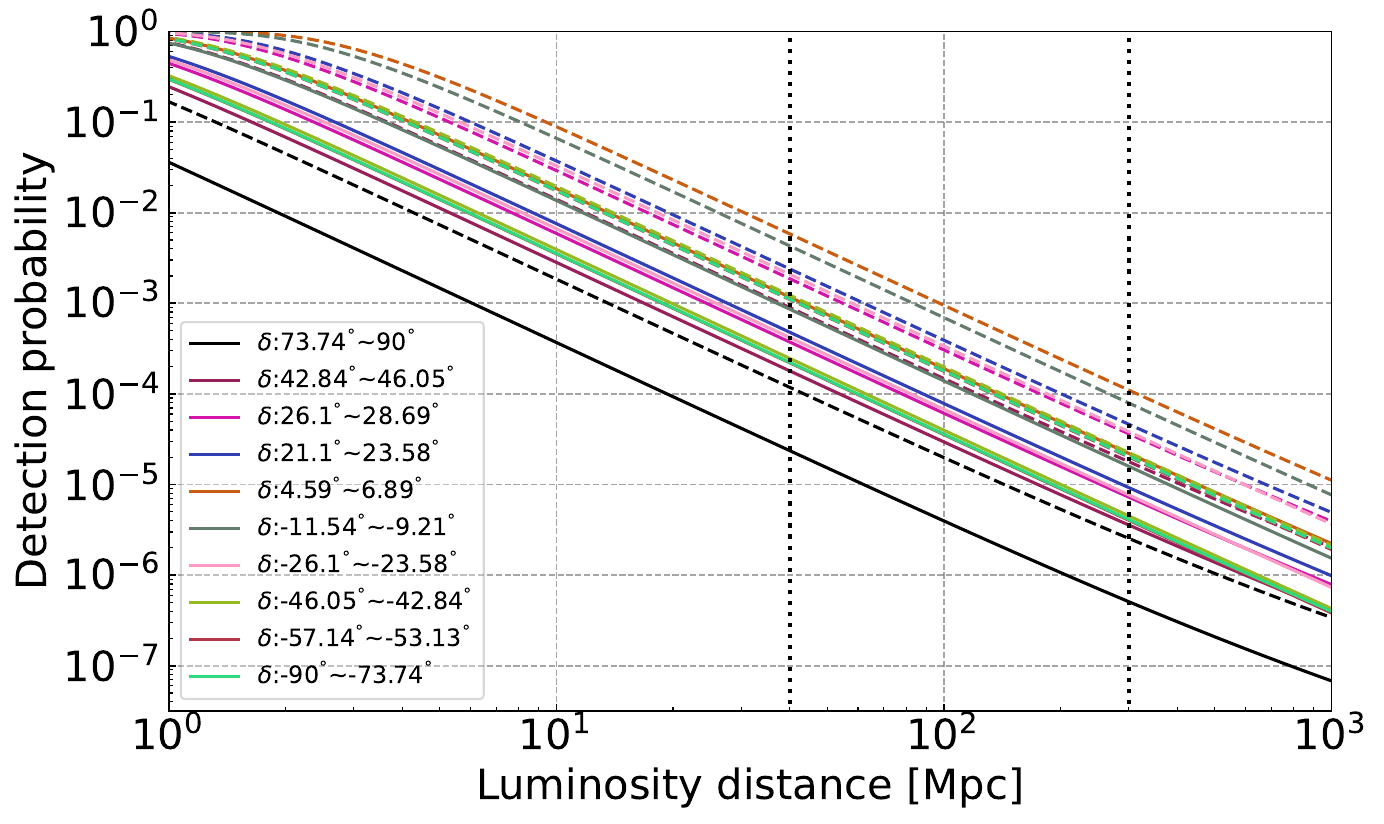}
      \caption{Similar to Figure \ref{Fig:7}, but for a GRB 230307A-like event with \(f_{p} = 0.7\) and \(\varepsilon _{p } \text{/} \varepsilon _{e } = 3\).}
      \label{Fig:9}
  \end{figure}

We investigate the detection prospects of high-energy neutrinos from GRB 230307A-like events which are from a Poynting-flux-dominated outflow, and the results are shown in Figure \ref{Fig:9}. Due to the inefficient neutrino production in the ICMART radius for GRB 230307A-like events, the detection of at least one high-energy neutrino is associated with GRB 230307A-like events. It requires that the distance of the source is below 3 Mpc even for IceCube-Gen2. This result indicates that the neutrinos from GRB 230307A-like events are not likely to be detected during the IceCube-Gen2 operation.

\begin{figure}[H]
      \centering
      \includegraphics[width=0.495\linewidth]{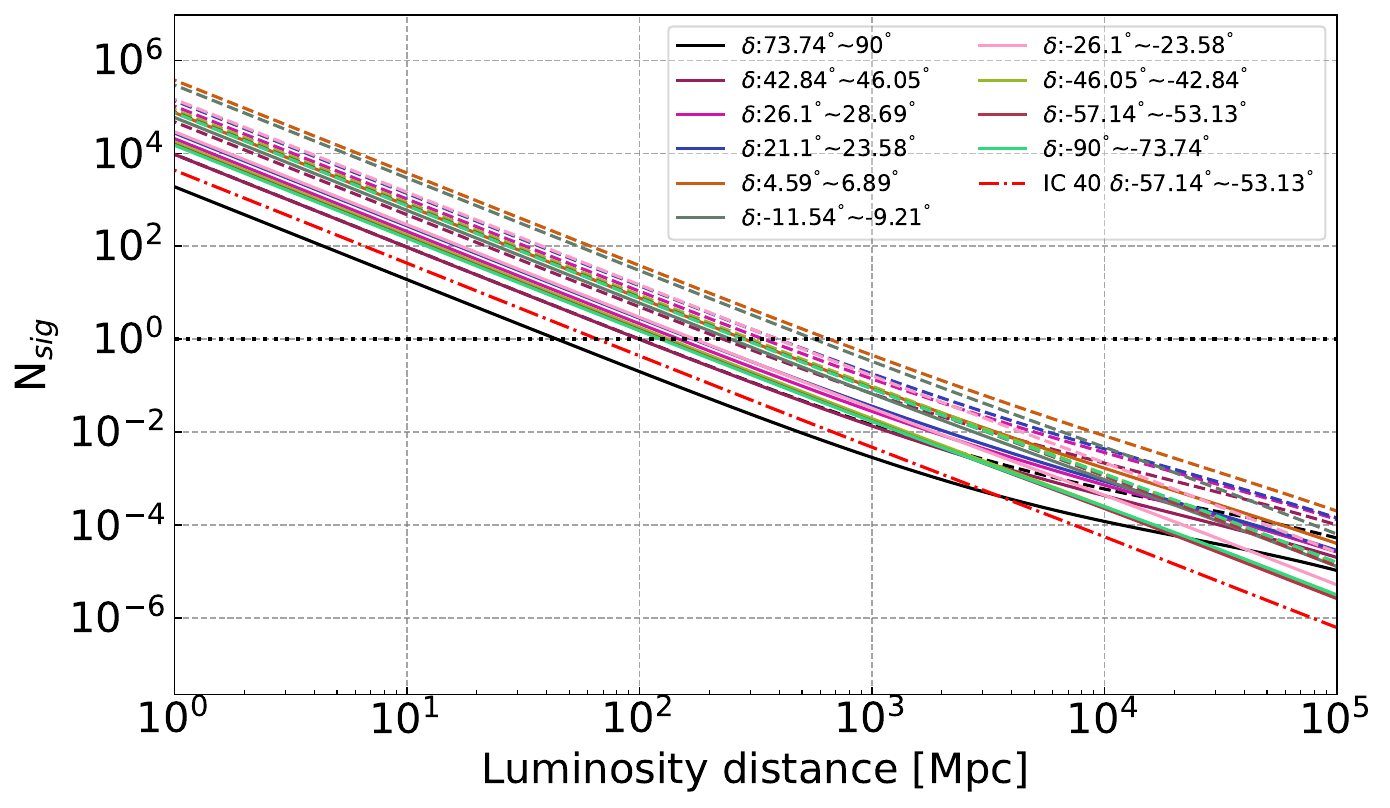}
      \includegraphics[width=0.495\linewidth]{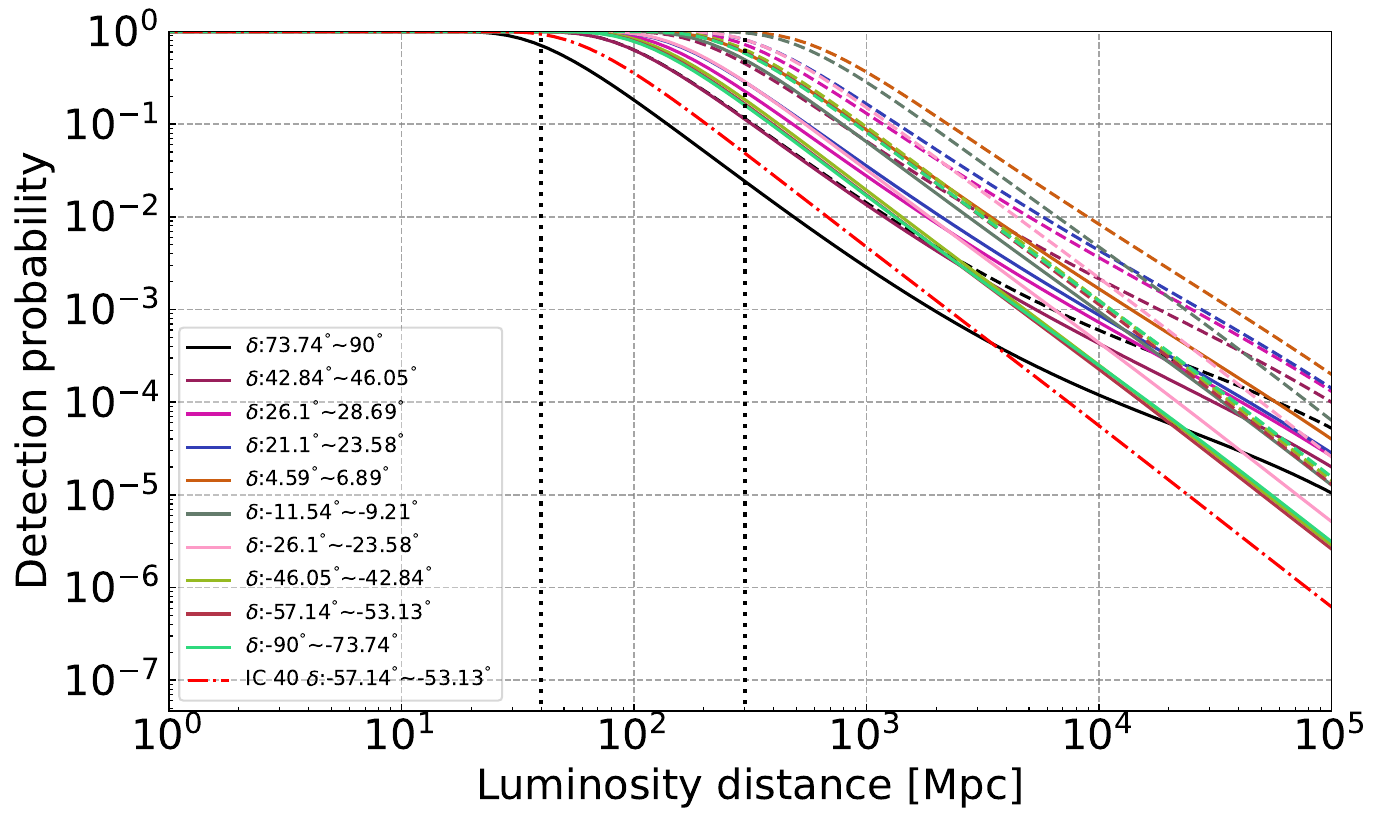}
      \caption{Similar to Figure \ref{Fig:9}, but for a GRB 080916C-like event.}
      \label{Fig:10}
  \end{figure}

Moreover, we also investigate the detection prospects of high-energy neutrinos from GRB 080916C-like events, and the results are shown in Figure \ref{Fig:10}. For an ideal case, the detection horizon of neutrinos from GRB 080916C-like events is 300 Mpc for IceCube, while one could potentially detect neutrinos from GRB 080916C-like events up to \(d_{L} \le 700\) Mpc for IceCube-Gen2. However, it is worth noting that if a GRB 080916C-like event with an isotropic energy of \(E_{\gamma} = 8.8\times10^{54}\text { erg}\) is located at 700 Mpc, it would be as bright as the GRB 221009A. If this is the case, the possibility of such an energetic event occurring in the nearby Universe is extremely low.

\section{Conclusion and discussion}\label{sec:5}
The new field of multimessenger astronomy is used to understand the main physical processes of astronomical sources in the Universe by adopting different types of messengers, such as electromagnetic radiation, gravitational waves, cosmic rays, and neutrinos. In this paper, based on the nonresults of neutrinos detected from IceCube Collaboration, we present constraints on physical parameters of four peculiar GRBs (080916C, 090902B, 211211A, 230307A) which are from two different types of jet composition (e.g., photosphere origin scenario of GRB 211211A and GRB 090902B, and Poynting-flux-dominated scenario of GRB 230307A and GRB 080916C). We find that only the dissipative photosphere model can be well constrained for these four peculiar GRBs. Within the scenario of photosphere origin, we obtain \(\varepsilon _{p } \text{/} \varepsilon _{e }<8\) for \(f_{p}>0.2\) from GRB 211211A and \(\varepsilon _{p } \text{/} \varepsilon _{e }<40\) for \(f_{p}>0.2\) from GRB 090902B. The UHECR hypothesis for GRB 211211A in the dissipative photosphere model could be ruled out. However, within the scenario of Poynting-flux-dominated scenario, we can effectively constrain neither GRB 230307A nor GRB 080916C. The nondetected neutrino from GRB 230307A is consistent with the predicted neutrino emission in the ICMART model, providing an independent clue to support the ICMART origin of GRB 230307A.

On the other hand, we also investigate the detection prospects of high-energy neutrinos from GRBs with two different types of jet composition. Within the scenario of photosphere origin, it is found that at least one high-energy neutrino associated with GRB 211211A-like events is possibly detected at \(d_{L} \le 700\) Mpc for IceCube-Gen2. It means that it is possible to detect a neutrino coincident with the GW signal if the GRB 211211A-like events originated from the merger of the compact stars with photosphere dissipation. For the bright GRB 090902B-like events with high redshift, it is possible to detect a neutrino at a distance greater than 1 Gpc. For the Poynting-flux-dominated scenario, the neutrinos from GRB 230307A-like events are not likely to be detected even during the IceCube-Gen2 operation. The detection horizon of neutrinos from GRB 080916C-like events is up to 700 Mpc for IceCube-Gen2. At such a distance, the event rate of GRB 080916C-like events may be extremely low. 

It is not surprising that the ICMART model is not well constrained. Given the low number density of photons at a large radius, it can lead to inefficient neutrino production. The composition of UHECRs in different energy ranges is quite different. For the high-energy range, the contribution from heavier nuclei will become more important \citep{2014PhRvD..90l2005A}, and the heavy nuclei are not likely to be photodisintegrated for such a low number density of photons at a large radius \citep{2008PhRvD..78b3005M}. If this is the case, the inefficient neutrino production is not caused by the inefficient UHECRs acceleration, but by the low number density of photons. Therefore, the GRBs that come from the ICMART model are the potential candidates for the accelerator of UHECR nuclei.

The upcoming neutrino detectors located near the equator and in the northern hemisphere, such as TRIDEN in the South China Sea and KM3NeT in the Mediterranean, will complete the lower sensitivity of IceCube in the low-energy region of the southern hemisphere. It will enhance our capability to detect neutrinos originating from the dissipative photosphere model, which predicts a lower energy of neutrinos.

\section*{Acknowledgments}
We thank Shun-Ke Ai and Xing Yang for the helpful discussion. This work is supported by the Guangxi Science Foundation of the National (grant No. 2023GXNSFDA026007), the Natural Science Foundation of China (grant Nos. 11922301 and 12133003), the Program of Bagui Scholars Program (LHJ), and the Guangxi Talent Program ("Highland of Innovation Talents").

\appendix

\section{Derivations of the break energy of neutrino spectrum}\label{App_a}
The photomeson interaction most likely proceeds at $\Delta$-resonance. The threshold condition for $\Delta$-resonance is
\begin{equation}
\begin{aligned}
E_{p} E_{\gamma} &\gtrsim \frac{m_{\Delta }^{2}-m_{p }^{2}  }{4}~\left(\frac{\Gamma}{1+z}\right)^{2} \\&=0.160~{\rm GeV^2}~\left(\frac{\Gamma}{1+z}\right)^{2}.
\label{eq:delta_condition}
\end{aligned}
\end{equation}
The $\pi^{+}$ typically share 1/5 of the proton energy, and each lepton produce by $\pi^{+}$ decay equally share 1/4 the parent $\pi^{+}$ energy. Thus the typical neutrino energy is
\begin{equation}
E_{\nu }\simeq0.05E_{p}.\label{eq:neutrino energy}
\end{equation}
Together with Equations (\ref{eq:delta_condition}) and (\ref{eq:neutrino energy}), we have the characteristic energy of neutrinos generated from a GRB given by
\begin{equation}
E_{\nu }  \simeq 0.008~{\rm GeV } ~\Gamma^{2} \left(1+z\right)^{-2}\left ( \frac{E_{\gamma}}{\text{GeV}} \right )^{-1},
\end{equation}
where $E_{\nu }$ and $E_{\gamma}$ are replaced by $\varepsilon _{\nu,1 }$ and $\varepsilon _{\gamma,b}$. Then, the characteristic value for Lorentz factor and $\varepsilon _{\gamma,b}$ is taken. This gives the low-energy break in the neutrino spectrum
\begin{equation}
     \begin{aligned}
     \varepsilon _{\nu,1 } = 7.2\times 10^{5}\text{GeV}\left ( \frac{\varepsilon _{\gamma,b}}{\text{MeV}} \right ) ^{-1}
\left ( \frac{\Gamma }{300} \right )^{2} \left ( 1+z \right ) ^{-2}.
     \end{aligned}
 \end{equation}

The high-energy break of the neutrino spectrum \(\varepsilon _{\nu,2 }\) is caused by the synchrotron cooling of charged pions because the time for charged pions to lose energy by synchrotron cooling is shorter than the time of decay if charged pions are energetic enough. The synchrotron cooling timescale of relativistic charged pions can be calculated as
\begin{equation}
t_{\pi^+,{\rm syn}}^{\prime} = \frac{6\pi m_{\pi^+} c}{\gamma_{\pi^+} \sigma_{T,\pi^+} B^{\prime 2}},
\end{equation}
where $\gamma_{\pi^+}$ is the Lorentz factor for the charged particle's random motion in the jet's comoving frame, $m_{\pi^+} = 0.15 m_p$ is the rest mass of $\pi^+$, $m_p = 1.67\times 10^{-24} \text{g}$ is the rest mass of proton. $\sigma_{T,\pi^+}$ is the Thomson scattering cross section of $\pi^+$ which can be obtained from the Thomson scattering cross section of the electron as $\sigma_{T,\pi^+} = (m_{e} / m_{\pi^+})^2 \sigma_{T,e}$, $\sigma_{T,e} = 6.65\times 10^{-25} \rm cm^{2}$ is the Thomson scattering cross section of the electron, $m_{e} = 9.1\times 10^{-28} \text{g}$ is the rest mass of the electron, \(B^{'} =\left [ \frac{L_{\rm GRB}\left ( \varepsilon _{B } \text{/} \varepsilon _{e }\right )}{2R^{2}\Gamma^{2}c }  \right ] ^{1/2}\) is the strength of the random magnetic field. The decay time of $\pi^{+}$ is
\begin{equation}
t_{\pi^{+},{\rm dec}}^{\prime} = \gamma_{\pi^+} \tau_{\pi^{+}},
\end{equation}
where $\tau_{\pi^{+}} = 2.8 \times 10^{-8}s$ is the decay time of $\pi^{+}$ at rest. $\gamma_{\pi^+}$ can be obtained by $t_{\pi^{+},{\rm dec}}^{\prime}$ equals to $t_{\pi^+,{\rm syn}}^{\prime}$. Since each lepton produced by $\pi^{+}$ decay equally shares 1/4 of the parent $\pi^{+}$ energy, the high-energy break of the neutrino spectrum is calculated as
\begin{equation}
     \begin{aligned}
       \varepsilon _{\nu,2 } =&\frac{1}{4} {\cal D} \gamma_{\pi^+} m_{\pi^+} c^2 \nonumber \\=&\text{ }1.17\times 10^{8}\text{GeV}L_{\rm GRB,52}^{-\frac{1}{2} } \varepsilon _{e,-1 }^{\frac{1}{2}}\\&\varepsilon _{B,-1 }^{-\frac{1}{2}}R_{13}\left ( \frac{\Gamma }{300} \right )^{2} \left ( 1+z \right ) ^{-1}.
     \end{aligned}
 \end{equation}
where ${\cal D} \approx 2\Gamma$ is the Doppler factor.

\bibliography{ms}
\bibliographystyle{aasjournal}



\end{document}